\pdfoutput=1

\documentclass[12pt,a4paper]{article}

\usepackage{ifthen} 
\newboolean{pdflatex}
\setboolean{pdflatex}{true} 

\newboolean{articletitles}
\setboolean{articletitles}{true} 

\newboolean{uprightparticles}
\setboolean{uprightparticles}{false} 

\def\paperauthors{LHCb collaboration} 
\def\paperasciititle{Measurement of CP observables in the process B0 to DK*0 with two- and four-body D decays} 
\def\papertitle{Measurement of \CP observables in the process $B^0 \to DK^{*0}$ with \\ two- and four-body $D$ decays} 
\def\paperkeywords{{High Energy Physics}, {LHCb}} 
\def\papercopyright{\the\year\ CERN for the benefit of the LHCb collaboration} 
\def\paperlicence{CC-BY-4.0 licence}
\def\paperlicenceurl{https://creativecommons.org/licenses/by/4.0/}

\usepackage[top=1in, bottom=1.25in, left=1in, right=1in]{geometry}

\usepackage{microtype}
\usepackage{lineno}  
\usepackage{xspace} 
\usepackage{caption} 

\usepackage{graphicx}  
\usepackage{color}
\usepackage{colortbl}
\graphicspath{{./figs/}} 
\DeclareGraphicsExtensions{.pdf,.PDF,png,.PNG}

\usepackage{amsmath} 
\usepackage{amssymb}
\usepackage{amsfonts}
\usepackage{upgreek} 

\newcommand*\patchAmsMathEnvironmentForLineno[1]{%
\expandafter\let\csname old#1\expandafter\endcsname\csname #1\endcsname
\expandafter\let\csname oldend#1\expandafter\endcsname\csname
end#1\endcsname
 \renewenvironment{#1}%
   {\linenomath\csname old#1\endcsname}%
   {\csname oldend#1\endcsname\endlinenomath}%
}
\newcommand*\patchBothAmsMathEnvironmentsForLineno[1]{%
  \patchAmsMathEnvironmentForLineno{#1}%
  \patchAmsMathEnvironmentForLineno{#1*}%
}
\AtBeginDocument{%
\patchBothAmsMathEnvironmentsForLineno{equation}%
\patchBothAmsMathEnvironmentsForLineno{align}%
\patchBothAmsMathEnvironmentsForLineno{flalign}%
\patchBothAmsMathEnvironmentsForLineno{alignat}%
\patchBothAmsMathEnvironmentsForLineno{gather}%
\patchBothAmsMathEnvironmentsForLineno{multline}%
\patchBothAmsMathEnvironmentsForLineno{eqnarray}%
}

\usepackage{hyperxmp}

\usepackage[pdftex,
            pdfauthor={\paperauthors},
            pdftitle={\paperasciititle},
            pdfkeywords={\paperkeywords},
            pdfcopyright={Copyright (C) \papercopyright},
            pdflicenseurl={\paperlicenceurl}]{hyperref}

\usepackage[colorinlistoftodos,textsize=scriptsize]{todonotes}

\usepackage[all]{hypcap} 

\usepackage{xspace} 
\usepackage{upgreek}

\def\lhcb   {\mbox{LHCb}\xspace}

\def\MagUp {\mbox{\em Mag\kern -0.05em Up}\xspace}

\ifthenelse{\boolean{uprightparticles}}%
{

 \def\Ppi         {\ensuremath{\uppi}\xspace}

 \def\PDelta      {\ensuremath{\Delta}\xspace}                 
 \def\PXi         {\ensuremath{\Xi}\xspace}                 
 \def\PLambda     {\ensuremath{\Lambda}\xspace}                 
 \def\PSigma      {\ensuremath{\Sigma}\xspace}                 
 \def\POmega      {\ensuremath{\Omega}\xspace}                 
 \def\PUpsilon    {\ensuremath{\Upsilon}\xspace}

 \def\PB      {\ensuremath{\mathrm{B}}\xspace}                 
                  
 \def\PD      {\ensuremath{\mathrm{D}}\xspace}

 \def\PK      {\ensuremath{\mathrm{K}}\xspace}

 \def\Pb      {\ensuremath{\mathrm{b}}\xspace}                 
 \def\Pc      {\ensuremath{\mathrm{c}}\xspace}                 
 \def\Pd      {\ensuremath{\mathrm{d}}\xspace}

 \def\Pi      {\ensuremath{\mathrm{i}}\xspace}

 \def\Ps      {\ensuremath{\mathrm{s}}\xspace}                 
 \def\Pt      {\ensuremath{\mathrm{t}}\xspace}                 
 \def\Pu      {\ensuremath{\mathrm{u}}\xspace}

 \def\thebaroffset{0.0em}
}
{

 \def\Ppi         {\ensuremath{\pi}\xspace}

 \mathchardef\PDelta="7101
 \mathchardef\PXi="7104
 \mathchardef\PLambda="7103
 \mathchardef\PSigma="7106
 \mathchardef\POmega="710A
 \mathchardef\PUpsilon="7107
                  
 \def\PB      {\ensuremath{B}\xspace}                 
                  
 \def\PD      {\ensuremath{D}\xspace}

 \def\PK      {\ensuremath{K}\xspace}

 \def\Pb      {\ensuremath{b}\xspace}                 
 \def\Pc      {\ensuremath{c}\xspace}                 
 \def\Pd      {\ensuremath{d}\xspace}

 \def\Pi      {\ensuremath{i}\xspace}

 \def\Ps      {\ensuremath{s}\xspace}                 
 \def\Pt      {\ensuremath{t}\xspace}                 
 \def\Pu      {\ensuremath{u}\xspace}

 \def\thebaroffset{0.18em}
}
\newcommand{\offsetoverline}[2][\thebaroffset]{\kern #1\overline{\kern -#1 #2}}%

\makeatletter
\ifcase \@ptsize \relax
  \newcommand{\miniscule}{\@setfontsize\miniscule{4}{5}}
\or
  \newcommand{\miniscule}{\@setfontsize\miniscule{5}{6}}
\or
  \newcommand{\miniscule}{\@setfontsize\miniscule{5}{6}}
\fi
\makeatother

\DeclareRobustCommand{\optbar}[1]{\shortstack{{\miniscule (\rule[.5ex]{1.25em}{.18mm})}
  \\ [-.7ex] $#1$}}

\def\uquark    {{\ensuremath{\Pu}}\xspace}

\def\dquark    {{\ensuremath{\Pd}}\xspace}

\def\squark    {{\ensuremath{\Ps}}\xspace}

\def\cquark    {{\ensuremath{\Pc}}\xspace}

\def\bquark    {{\ensuremath{\Pb}}\xspace}

\def\tquark    {{\ensuremath{\Pt}}\xspace}

\def\pion   {{\ensuremath{\Ppi}}\xspace}
\def\piz    {{\ensuremath{\pion^0}}\xspace}
\def\pip    {{\ensuremath{\pion^+}}\xspace}
\def\pim    {{\ensuremath{\pion^-}}\xspace}

\def\pimp   {{\ensuremath{\pion^\mp}}\xspace}

\def\kaon    {{\ensuremath{\PK}}\xspace}
\def\Kbar    {{\ensuremath{\offsetoverline{\PK}}}\xspace}

\def\KorKbar {\kern \thebaroffset\optbar{\kern -\thebaroffset \PK}{}\xspace}
\def\Kz      {{\ensuremath{\kaon^0}}\xspace}
\def\Kzb     {{\ensuremath{\Kbar{}^0}}\xspace}
\def\Kp      {{\ensuremath{\kaon^+}}\xspace}
\def\Km      {{\ensuremath{\kaon^-}}\xspace}
\def\Kpm     {{\ensuremath{\kaon^\pm}}\xspace}

\def\KS      {{\ensuremath{\kaon^0_{\mathrm{S}}}}\xspace}

\def\Kstarz  {{\ensuremath{\kaon^{*0}}}\xspace}
\def\Kstarzb {{\ensuremath{\Kbar{}^{*0}}}\xspace}
\def\Kstar   {{\ensuremath{\kaon^*}}\xspace}

\def\Dbar    {{\ensuremath{\offsetoverline{\PD}}}\xspace}
\def\D       {{\ensuremath{\PD}}\xspace}

\def\DorDbar {\kern \thebaroffset\optbar{\kern -\thebaroffset \PD}\xspace}
\def\Dz      {{\ensuremath{\D^0}}\xspace}
\def\Dzb     {{\ensuremath{\Dbar{}^0}}\xspace}
\def\Dp      {{\ensuremath{\D^+}}\xspace}
\def\Dm      {{\ensuremath{\D^-}}\xspace}

\def\Dstar   {{\ensuremath{\D^*}}\xspace}

\def\Dstarz  {{\ensuremath{\D^{*0}}}\xspace}
\def\Dstarzb {{\ensuremath{\Dbar{}^{*0}}}\xspace}

\def\Ds      {{\ensuremath{\D^+_\squark}}\xspace}

\def\Dsm     {{\ensuremath{\D^-_\squark}}\xspace}

\def\B       {{\ensuremath{\PB}}\xspace}
\def\Bbar    {{\ensuremath{\offsetoverline{\PB}}}\xspace}

\def\BorBbar {\kern \thebaroffset\optbar{\kern -\thebaroffset \PB}\xspace}
\def\Bz      {{\ensuremath{\B^0}}\xspace}
\def\Bzb     {{\ensuremath{\Bbar{}^0}}\xspace}
\def\Bd      {{\ensuremath{\B^0}}\xspace}

\def\BdorBdbar {\kern \thebaroffset\optbar{\kern -\thebaroffset \Bd}\xspace}
\def\Bu      {{\ensuremath{\B^+}}\xspace}

\def\Bp      {{\ensuremath{\Bu}}\xspace}

\def\Bs      {{\ensuremath{\B^0_\squark}}\xspace}
\def\Bsb     {{\ensuremath{\Bbar{}^0_\squark}}\xspace}
\def\BsorBsbar {\kern \thebaroffset\optbar{\kern -\thebaroffset \Bs}\xspace}

\def\Y#1S{\ensuremath{\PUpsilon{(#1S)}}\xspace}

\def\LorLbar     {\kern \thebaroffset\optbar{\kern -\thebaroffset \PLambda}\xspace}

\def\BF         {{\ensuremath{\mathcal{B}}}\xspace}

\def\to                 {\ensuremath{\rightarrow}\xspace}

\def\CP                {{\ensuremath{C\!P}}\xspace}

\def\Vud  {{\ensuremath{V_{\uquark\dquark}}}\xspace}
\def\Vcd  {{\ensuremath{V_{\cquark\dquark}}}\xspace}
\def\Vtd  {{\ensuremath{V_{\tquark\dquark}}}\xspace}
\def\Vus  {{\ensuremath{V_{\uquark\squark}}}\xspace}
\def\Vcs  {{\ensuremath{V_{\cquark\squark}}}\xspace}

\def\Vub  {{\ensuremath{V_{\uquark\bquark}}}\xspace}
\def\Vcb  {{\ensuremath{V_{\cquark\bquark}}}\xspace}
\def\Vtb  {{\ensuremath{V_{\tquark\bquark}}}\xspace}

\def\AT#1     {\ensuremath{A_{\mathrm{T}}^{#1}}\xspace}           

\def\C#1      {\ensuremath{\mathcal{C}_{#1}}\xspace}                       
\def\Cp#1     {\ensuremath{\mathcal{C}_{#1}^{'}}\xspace}                    
\def\Ceff#1   {\ensuremath{\mathcal{C}_{#1}^{\mathrm{(eff)}}}\xspace}        
\def\Cpeff#1  {\ensuremath{\mathcal{C}_{#1}^{'\mathrm{(eff)}}}\xspace}       
\def\Ope#1    {\ensuremath{\mathcal{O}_{#1}}\xspace}                       
\def\Opep#1   {\ensuremath{\mathcal{O}_{#1}^{'}}\xspace}                    

\newcommand{\nospaceunit}[1]{\ensuremath{\text{#1}}}       
\newcommand{\aunit}[1]{\ensuremath{\text{\,#1}}}       

\newcommand{\tev}{\aunit{Te\kern -0.1em V}\xspace}
\newcommand{\gev}{\aunit{Ge\kern -0.1em V}\xspace}
\newcommand{\mev}{\aunit{Me\kern -0.1em V}\xspace}
\newcommand{\kev}{\aunit{ke\kern -0.1em V}\xspace}
\newcommand{\ev}{\aunit{e\kern -0.1em V}\xspace}
\newcommand{\mevc}{\ensuremath{\aunit{Me\kern -0.1em V\!/}c}\xspace}
\newcommand{\gevc}{\ensuremath{\aunit{Ge\kern -0.1em V\!/}c}\xspace}
\newcommand{\mevcc}{\ensuremath{\aunit{Me\kern -0.1em V\!/}c^2}\xspace}
\newcommand{\gevcc}{\ensuremath{\aunit{Ge\kern -0.1em V\!/}c^2}\xspace}

\def\mm   {\aunit{mm}\xspace}

\def\mum  {\ensuremath{\,\upmu\nospaceunit{m}}\xspace}

\def\fb   {\ensuremath{\aunit{fb}}\xspace}
\def\invfb   {\ensuremath{\fb^{-1}}\xspace}

\def\ps   {\ensuremath{\aunit{ps}}\xspace}

\newcommand{\chisq}{\ensuremath{\chi^2}\xspace}
\newcommand{\chisqndf}{\ensuremath{\chi^2/\mathrm{ndf}}\xspace}
\newcommand{\chisqip}{\ensuremath{\chi^2_{\text{IP}}}\xspace}

\def\gsim{{~\raise.15em\hbox{$>$}\kern-.85em
          \lower.35em\hbox{$\sim$}~}\xspace}
\def\lsim{{~\raise.15em\hbox{$<$}\kern-.85em
          \lower.35em\hbox{$\sim$}~}\xspace}

\def\sqs   {\ensuremath{\protect\sqrt{s}}\xspace}

\def\pt         {\ensuremath{p_{\mathrm{T}}}\xspace}

\def\ptot       {\ensuremath{p}\xspace}

\def\evtgen     {\mbox{\textsc{EvtGen}}\xspace}

\def\geant      {\mbox{\textsc{Geant4}}\xspace}

\def\photos     {\mbox{\textsc{Photos}}\xspace}

\def\pythia     {\mbox{\textsc{Pythia}}\xspace}

\xspace

\def\tell1  {TELL1\xspace}
\def\ukl1   {UKL1\xspace}

\usepackage{cite} 
\usepackage{mciteplus}

\usepackage{placeins}
\usepackage{booktabs}
\usepackage{rotating}
\usepackage{longtable} 
\usepackage{gensymb}
\usepackage{amsmath}
\usepackage[bottom]{footmisc}

\begin{document}

\renewcommand{\thefootnote}{\fnsymbol{footnote}}
\setcounter{footnote}{1}


\begin{titlepage}
\pagenumbering{roman}

\vspace*{-1.5cm}
\centerline{\large EUROPEAN ORGANIZATION FOR NUCLEAR RESEARCH (CERN)}
\vspace*{1.5cm}
\noindent
\begin{tabular*}{\linewidth}{lc@{\extracolsep{\fill}}r@{\extracolsep{0pt}}}
\ifthenelse{\boolean{pdflatex}}
{\vspace*{-1.5cm}\mbox{\!\!\!\includegraphics[width=.14\textwidth]{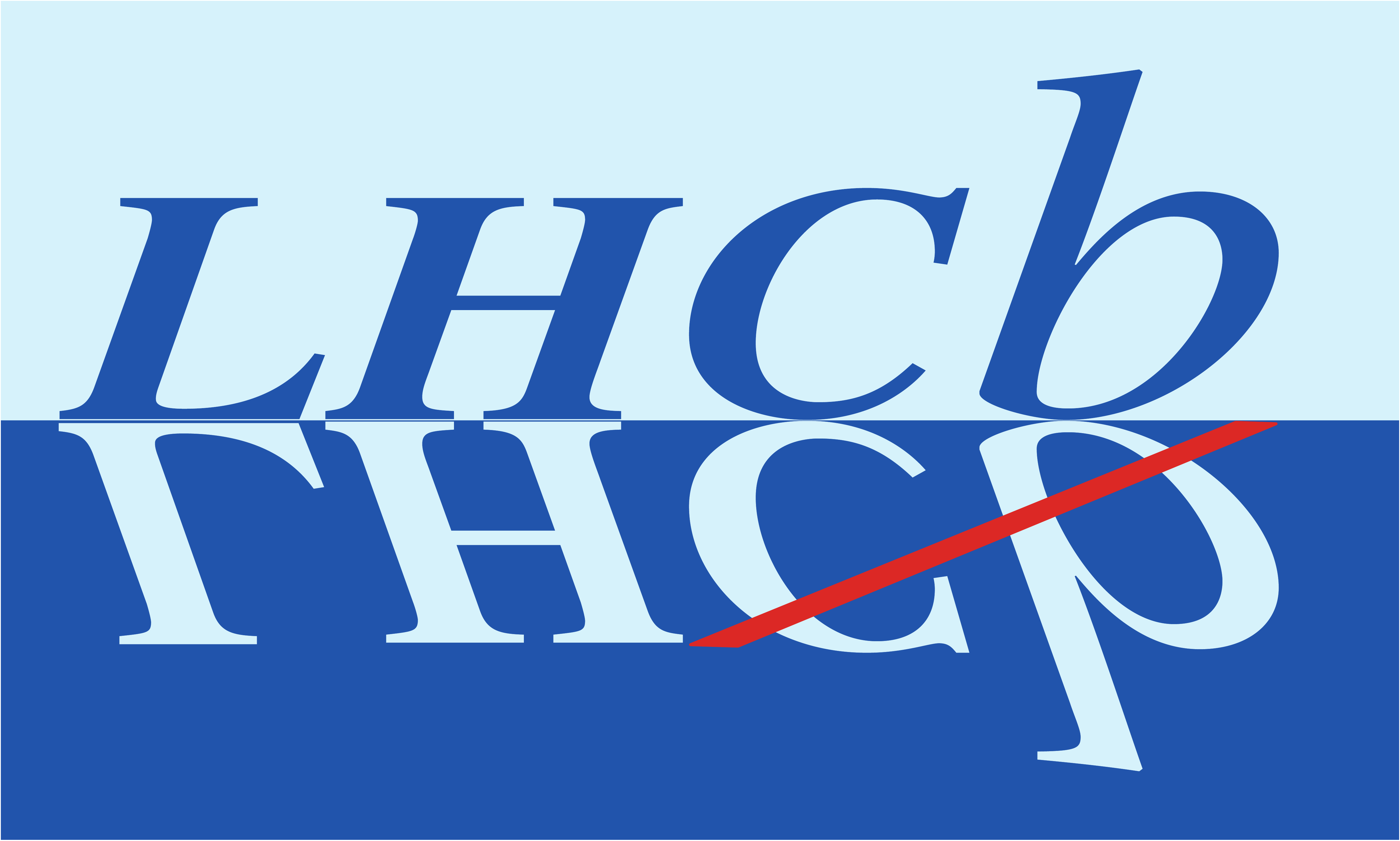}} & &}%
{\vspace*{-1.2cm}\mbox{\!\!\!\includegraphics[width=.12\textwidth]{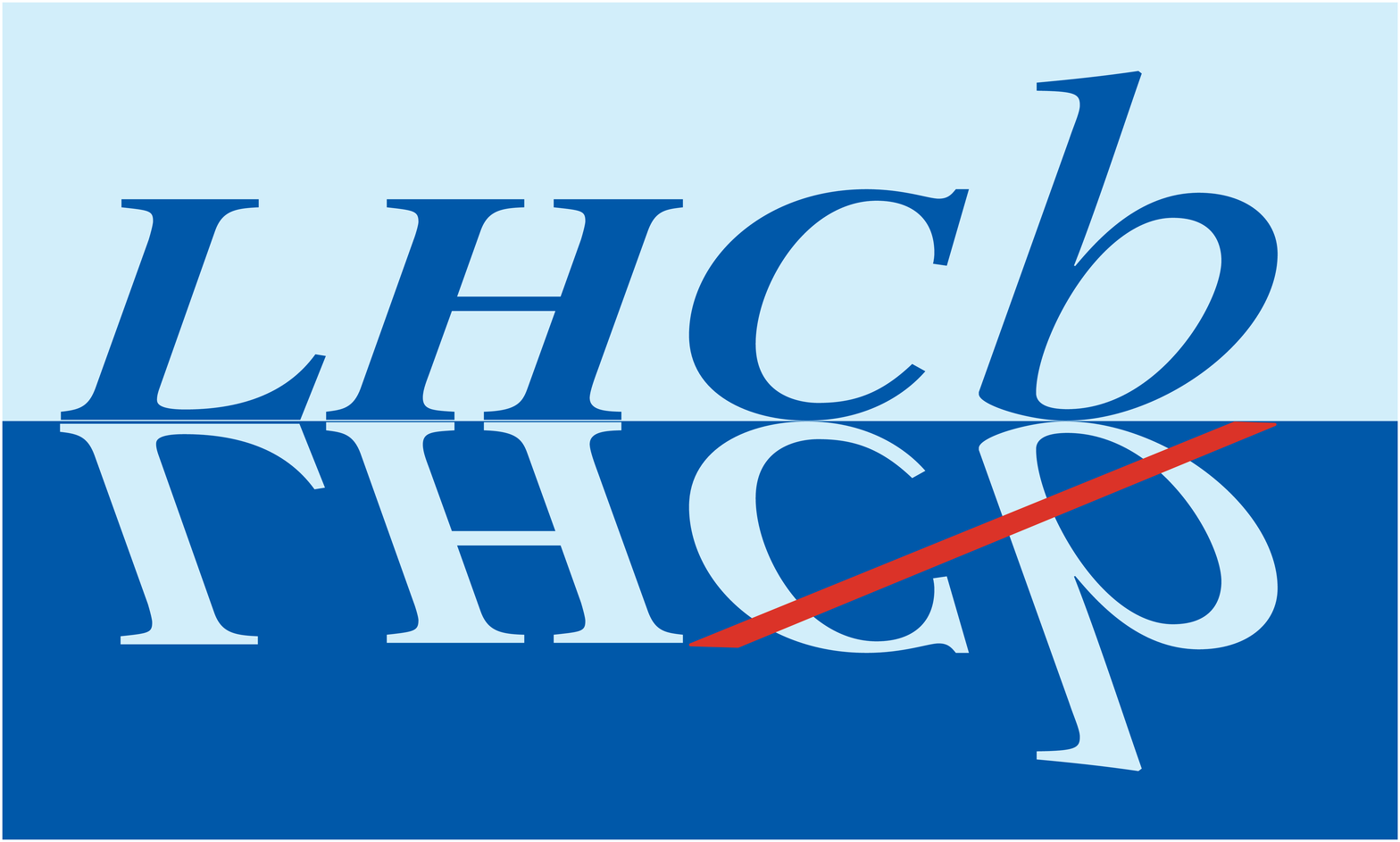}} & &}%
\\
 & & CERN-EP-2019-111 \\
 & & LHCb-PAPER-2019-021 \\
 & & August 13, 2019 \\ 
 & & \\
\end{tabular*}

\vspace*{4.0cm}

{\normalfont\bfseries\boldmath\huge
\begin{center}
  \papertitle 
\end{center}
}

\vspace*{2.0cm}

\begin{center}
\paperauthors\footnote{Authors are listed at the end of this paper.}
\end{center}

\vspace{\fill}

\begin{abstract}
  \noindent
    Measurements of \CP\ observables in $\Bz \to D\Kstarz$ decays are presented, where $D$ represents a superposition of $\Dz$ and $\Dzb$ states. The $D$ meson is reconstructed in the two-body final states $\Kp\pim$, $\pip\Km$, $\Kp\Km$ and $\pip\pim$, and, for the first time, in the four-body final states $\Kp\pim\pip\pim$, $\pip\Km\pip\pim$ and $\pip\pim\pip\pim$. The analysis uses a sample of neutral $B$ mesons produced in proton-proton collisions, corresponding to an integrated luminosity of 1.0, 2.0 and 1.8\invfb\ collected with the \lhcb\ detector at centre-of-mass energies of $\sqs = $ 7, 8 and 13\tev, respectively.   First observations of the decays $\Bz \to D(\pip\Km)\Kstarz$ and $\Bz \to D(\pip\pim\pip\pim)\Kstarz$ are obtained.  The measured observables are interpreted in terms of the $\CP$-violating weak phase $\gamma$.
\end{abstract}

\vspace*{2.0cm}

\begin{center}
  Published in JHEP 08 (2019) 041
\end{center}

\vspace{\fill}

{\footnotesize 
\centerline{\copyright~\papercopyright. \href{\paperlicenceurl}{\paperlicence}.}}
\vspace*{2mm}

\end{titlepage}


\newpage
\setcounter{page}{2}
\mbox{~}
%
%
%
%

\cleardoublepage


\renewcommand{\thefootnote}{\arabic{footnote}}
\setcounter{footnote}{0}



\pagestyle{plain} 
\setcounter{page}{1}
\pagenumbering{arabic}


\section{Introduction}
\label{sec:Introduction}

In the Standard Model, \CP\ violation is described by the irreducible complex phase of the Cabibbo--Kobayashi--Maskawa (CKM) quark mixing matrix\cite{Cabibbo:1963yz,Kobayashi:1973fv}. This matrix is unitary, leading to the condition $\Vud\Vub^* + \Vcd\Vcb^* + \Vtd\Vtb^* = 0$, where $V_{ij}$ is the CKM matrix element relating quark $i$ to quark $j$. This relation can be represented as a triangle in the complex plane, with angles $\alpha$, $\beta$ and $\gamma$. Improving knowledge of $\gamma$ is one of the most important goals in flavour physics.  This angle is defined as $\gamma \equiv \arg{(-{\Vud\Vub^*}/{\Vcd\Vcb^*})}$, which is equal to $\arg{(-{\Vus\Vub^*}/{\Vcs\Vcb^*})}$ up to $\mathcal{O}(\lambda^4) \sim 10^{-3}$\cite{wolfenstein}. This can be measured through the interference of  $b \to c$ and $b \to u$ transition amplitudes in tree-level $b$-hadron decays.\footnote{Except where stated otherwise, the inclusion of charge-conjugate processes is implied throughout this paper.} Such a measurement provides a Standard-Model benchmark against which observables determined in loop-mediated processes, expected to be more susceptible to the influence of physics beyond the Standard Model, can be compared.

Measurements from the LHCb experiment yield $\gamma = (74.0\,^{+5.0}_{-5.8})$\degree~\cite{LHCb-CONF-2018-002,LHCb-PAPER-2016-032}, which is the most precise determination of $\gamma$ from a single experiment.  The precision is dominated by measurements exploiting the decay $\Bp \to D \Kp$, where $D$ indicates a superposition of $\Dz$ and $\Dzb$ mesons reconstructed in a final state common to both.  In order to test internal consistency, and to improve overall sensitivity, it is important to complement these measurements with those based on other decay modes. One important example is $\Bz \to D \Kstarz$ \cite{Dunietz:1991yd}, where $\Kstarz$ is the $\Kstar(892)^0$ meson and is reconstructed in its decay to $\Kp\pim$. This process involves the interference of $\Bz \to \Dzb\Kstarz$ decays, which proceed via a $b\to c$ quark transition, and $\Bz \to \Dz\Kstarz$decays, which involve a $b \to u$ quark transition and are therefore suppressed relative to $\Bz \to \Dzb\Kstarz$.  Feynman diagrams of these decays are shown in Fig.~\ref{fig:feynman}. Both transitions are colour-suppressed, in contrast to the charged $B$-meson case where only the $b \to u$ transition is colour-suppressed.  This leads to a greater suppression of the overall decay rates, but with the benefit of enhanced interference effects with respect to $\Bp \to D\Kp$ decays. The ratio $r_B^{D\!\Kstarz}$ between the magnitudes of the suppressed and favoured $\Bz$ decay amplitudes is expected to be around three times larger than the corresponding parameter in $\Bp \to D\Kp$ decays.

\FloatBarrier\begin{figure}
    \centering
    \includegraphics[width=0.9\textwidth]{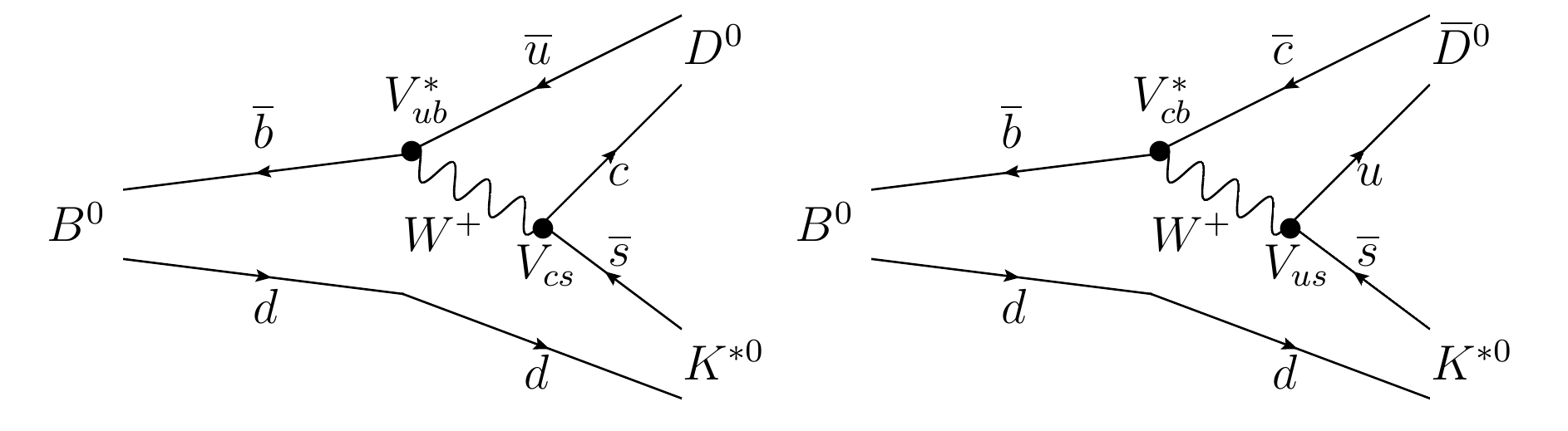}
    \caption{Feynman diagrams of (left) $\Bz \to \Dz\Kstarz$ and (right) $\Bz \to \Dzb\Kstarz$.}
\end{figure}
\label{fig:feynman}

The LHCb collaboration has performed studies of $\Bz \to D \Kstarz$ decays using data corresponding to an integrated luminosity of 3.0\invfb, reconstructing the $D$ meson in the two-body final states $\Kpm\pi^\mp$, $\Kp\Km$ and $\pip\pim$~\cite{LHCb-PAPER-2014-028}, and also the self-conjugate modes $\KS\pip\pim$ and $\KS\Kp\Km$~\cite{LHCb-PAPER-2016-006,LHCb-PAPER-2016-007}. In addition, the two-body $D$ decay modes $\Kp\pim$, $\Kp\Km$ and $\pip\pim$ have previously been exploited in an amplitude analysis of $\Bz \to D \Kp \pim$ decays, including $\Bz \to D\Kstarz$ decays~\cite{LHCb-PAPER-2015-059}.

In this paper, results are presented for a study of $\Bz \to D \Kstarz$ decays performed on a data set corresponding to 3.0\invfb\ of integrated luminosity collected at centre-of-mass energies of 7 and 8\tev\ during Run 1 of the LHC, and 1.8\invfb\ collected at 13\tev\ during Run 2 in 2015 and 2016. Observables sensitive to $\gamma$ are measured for the following final states of the $D$-meson decay:  $\Kpm\pimp$, $\Kp\Km$, $\pip\pim$, $\Kpm\pimp\pip\pim$ and $\pip\pim\pip\pim$. The study of the two-body modes benefits from several improvements 
with respect to Ref.~\cite{LHCb-PAPER-2014-028}, as well as from the larger data set.  The four-body modes are analysed for the first time in this decay chain.  The measurements involving $D \to \pip\pim\pip\pim$ are based on Run 2 data alone, 
as the central processing that performs the first step of the selection did not include a suitable selection for this mode in Run 1.

The paper is organised as follows:  Section~\ref{sec:observables} presents the observables to be measured, and their relationships to the physics parameters of interest;  Sect.~\ref{sec:Detector} discusses those aspects of the detector, trigger and simulation that are relevant for the measurement;  Sects.~\ref{sec:selection},~\ref{sec:mass_fit} and~\ref{sec:systematics} describe the candidate selection, the fit of the mass spectra and the assignment of systematic uncertainties, respectively;  the results, and their interpretation, are given in Sect.~\ref{sec:results}; and conclusions are presented in Sect.~\ref{sec:conclusion}.

\section{Analysis strategy}
\label{sec:observables}

This analysis exploits the interference between $\Bz \to \Dzb\Kstarz$ and $\Bz \to \Dz\Kstarz$ decays, with the $\Dz$ and $\Dzb$ mesons reconstructed in a common final state. The partial widths of these decays are used to construct observables, which have a dependence on $\gamma$ and the following parameters: the ratio $r_B^{D\!\Kstarz}$ between the magnitudes of the suppressed and favoured $\Bz$ decay amplitudes; the \CP-conserving strong-phase difference $\delta_B^{D\!\Kstarz}$ between the amplitudes; and a coherence factor $\kappa$, which accounts for other amplitudes that may contribute to the $\Bz \to D \Kp \pim$ final state in addition to the two diagrams responsible for the $\Bz \to D \Kstarz$ signal process.  Detailed definitions of these parameters may be found in Ref.~\cite{LHCb-PAPER-2014-028}.  An amplitude analysis of $\Bz \to D\Kp\pim$ decays has determined the coherence factor to be $\kappa = 0.958\,^{+0.005}_{-0.046}$ for the $\Kstarz$ selection criteria used in this measurement (see Sect.~\ref{sec:selection})~\cite{LHCb-PAPER-2015-059}, which indicates an almost pure $D\Kstarz$ sample.

Reconstructing the charmed meson through a decay to a \CP\ eigenstate, such as 
$D \to \Kp\Km$ or $D \to \pip\pim$, brings information on $\gamma$ through a strategy first proposed by Gronau, London and Wyler (GLW)~\cite{GLW1,GLW2}.  The asymmetry 
\begin{equation}
    \mathcal{A}_{\CP} \equiv \frac{\Gamma(\Bzb \to D_{\CP}\Kstarzb) - \Gamma(\Bz \to D_{\CP}\Kstarz)}{\Gamma(\Bzb \to D_{\CP}\Kstarzb) + \Gamma(\Bz \to D_{\CP}\Kstarz)}{\rm ,}
\label{eq:A_hh}
\end{equation}
where $\Gamma$ represents a partial decay width, is measured for both modes, yielding $\mathcal{A}^{KK}_{\CP}$ and $\mathcal{A}^{\pi\pi}_{\CP}$, which are expected to be equal when the small \CP-violating effects observed in the $D$-meson decay~\cite{LHCb-PAPER-2019-006} are neglected; this assumption applies for the remainder of the discussion. The asymmetry is related to the underlying parameters through
\begin{equation}
\label{eq:acpdep}
    \mathcal{A}_{\CP} = \frac{2 \kappa r^{D\!\Kstarz}_B \sin\delta^{D\!\Kstarz}_B \sin\gamma}{\mathcal{R}_{\CP}},
\end{equation}

\noindent where $\mathcal{R}_{\CP}$ is the charge-averaged rate of decays involving a $D$ meson decaying to a \CP\ eigenstate, defined as 
\begin{equation}
    {\mathcal R}_{\CP} \equiv 2\frac{\Gamma(\Bzb \to D_{\CP}\Kstarzb) +
        \Gamma(\Bz \to D_{\CP}\Kstarz)}{\Gamma(\Bzb \to \Dz\Kstarzb) + \Gamma(\Bz \to \Dzb \Kstarz)}.
\end{equation}
\noindent This is related to $\gamma$ and the auxiliary parameters through
\begin{equation}
\label{eq:rcpdep}
    \mathcal{R}_{\CP} = 1 + {(r^{D\!\Kstarz}_B)}^2 + 2 \kappa r^{D\!\Kstarz}_B \cos\delta^{D\!\Kstarz}_B \cos\gamma.
\end{equation}
Experimentally it is convenient to access $\mathcal{R}_{\CP}$ by noting that it is closely approximated by
\begin{equation}
\label{eq:rhhcp}
        \mathcal{R}_{\CP}^{hh} \equiv \frac{\Gamma(\Bzb \to D(h^+h^-)\Kstarzb) +
        \Gamma(\Bz \to D(h^+h^-)\Kstarz)}{\Gamma(\Bzb \to D(\Km\pip)\Kstarzb) + \Gamma(\Bz \to D(\Kp\pim)\Kstarz)}
        \times \frac{\BF(D^0 \to \Km\pip)}{\BF(D^0 \to h^+h^-)}\mathrm{,}
\end{equation}
where the branching fractions $\BF$ are known\cite{PDG2018}.

As proposed in Refs.~\cite{Nayak:2014tea,Malde:2015mha}, multibody $D$-meson decays to self-conjugate final states may be used in a quasi-GLW analysis provided their fractional \CP content is known.  Hence the observables ${\mathcal A}^{4\pi}_{\CP}$ and ${\mathcal R}^{4\pi}_{\CP}$ are measured, which are analogous to the two-body observables $\mathcal{A}_{\CP}$ and $\mathcal{R}^{hh}_{\CP}$, but for the decay $D\to \pip\pim\pip\pim$.  These new observables can be interpreted through equivalent expressions to Eqs.~\ref{eq:acpdep} and~\ref{eq:rcpdep} in which the interference terms acquire a factor of 
$(2F_+^{4\pi}-1)$, where $F_+^{4\pi}$ is the fractional \CP-even content of the decay, measured to be $0.769\pm 0.023$ from quantum-correlated $D$-meson decays~\cite{Harnew:2017tlp}.

The decays $D \to \Kpm\pimp$ are exploited in a method proposed by Atwood, Dunietz and Soni (ADS)~\cite{ADS0,ADS}. 
Considering the decays $\Kstarz \to \Kp\pim$ and $\Kstarzb \to \Km\pip$, four categories are defined: two decays with the same charge of the final-state kaons, which are favoured and labelled $K\pi$, and two decays with the opposite charge of the final-state kaons, which are suppressed and labelled $\pi K$. The interference effects, and hence sensitivity to $\gamma$, are expected to be substantial for the suppressed modes, and smaller for the favoured modes.

The partial-rate asymmetry of the suppressed ADS decays is given by
\begin{equation}
\label{eq:AADS}
{\mathcal A}^{\pi K}_{\rm ADS} \equiv
\frac{\Gamma(\Bzb \to D(\pim \Kp)\Kstarzb) - \Gamma(\Bz \to D(\pip \Km)\Kstarz)}{\Gamma(\Bzb \to D(\pim \Kp)\Kstarzb) + \Gamma(\Bz \to D(\pip \Km)\Kstarz)},
\end{equation}
and the charge-averaged rate with respect to the favoured modes by
\begin{equation}
\label{eq:RADS}
{\mathcal R}^{\pi K}_{\rm ADS} \equiv
\frac{\Gamma(\Bzb \to D(\pim \Kp)\Kstarzb) + \Gamma(\Bz \to D(\pip \Km)\Kstarz)}{\Gamma(\Bzb \to D(\Km \pip)\Kstarzb) + \Gamma(\Bz \to D(\Kp\pim)\Kstarz)},
\end{equation}
which have the following dependence on $\gamma$ and the auxiliary parameters:
\begin{equation}
\label{eq:AADS_dep}
    {\mathcal A}^{\pi K}_{\rm ADS} = \frac{2 \kappa r_B^{D\!\Kstarz}r_D^{K\pi}\sin(\delta_B^{D\!\Kstarz}+\delta_D^{K\pi})\sin\gamma}
    {(r_B^{D\!\Kstarz})^2 + (r_D^{K\pi})^2 + 2 \kappa r_B^{D\!\Kstarz}r_D^{K\pi}\cos(\delta_B^{D\!\Kstarz}+\delta_D^{K\pi})\cos\gamma},
\end{equation}
\begin{equation}
\label{eq:RADS_dep}
{\mathcal R}^{\pi K}_{\rm ADS} = \frac{(r_B^{D\!\Kstarz})^2 + (r_D^{K\pi})^2 + 2\kappa r_B^{D\!\Kstarz}r_D^{K\pi}\cos(\delta_B^{D\!\Kstarz} + \delta_D^{K\pi}) \cos\gamma}
    {1 + (r_B^{D\!\Kstarz} r_D^{K\pi})^2 + 2\kappa r_B^{D\!\Kstarz}r_D^{K\pi}\cos(\delta_B^{D\!\Kstarz} + \delta_D^{K\pi}) \cos\gamma}.
\end{equation}
Here, $r_D^{K\pi}=0.059\pm0.001$ is the ratio between the doubly Cabibbo-suppressed and Cabibbo-favoured decay amplitudes of the neutral charm meson, and $\delta_D^{K\pi} = (192.1\,^{+\,8.6}_{-10.2})^\circ$ is a strong-phase difference between the amplitudes~\cite{HFLAV16}.\footnote{All expressions and charm strong-phase values are given in the convention $\CP |\Dz\rangle = |\Dzb\rangle$. This implies a 180\degree\ offset with respect to the values quoted in Ref.~\cite{HFLAV16}, which are defined with a different sign convention.}  

The quantities measured experimentally are the ratios

\begin{equation}
\label{eq:Rplus}
    \mathcal{R}_+^{\pi K} = \frac{\Gamma(\Bz \to D(\pip \Km)\Kstarz)}{\Gamma(\Bz \to D(\Kp\pim)\Kstarz)}
\end{equation}
\noindent and
\begin{equation}
\label{eq:Rminus}
    \mathcal{R}_-^{\pi K} = \frac{\Gamma(\Bzb \to D(\pim \Kp)\Kstarzb)}{\Gamma(\Bzb \to D(\Km\pip)\Kstarzb)}.
\end{equation}

\noindent The relationships 
\begin{equation}
     {\mathcal A}^{\pi K}_{\rm ADS} \simeq (\mathcal{R}^{\pi K}_{-} - \mathcal{R}^{\pi K}_{+})/ (\mathcal{R}^{\pi K}_{-} + \mathcal{R}^{\pi K}_{+})
\end{equation}
 and 
\begin{equation}
 {\mathcal R}^{\pi K}_{\rm ADS} \simeq (\mathcal{R}^{\pi K}_{+} + \mathcal{R}^{\pi K}_{-}) /2
\end{equation}
 allow the ADS observables to be recovered, where the approximate equalities are exact in the absence of \CP\ asymmetry in the favoured modes.

The ADS method can be extended in an analogous way to the four-body mode $D\to K^\pm\pi^\mp\pip\pim$,  with observables $\mathcal{R}_{\pm}^{\pi K \pi\pi}$.  In interpreting the results it is necessary to account for the variation of amplitude across the phase space of the $D$-meson decay. 
In the equivalent relations for Eqs.~\ref{eq:AADS_dep} and~\ref{eq:RADS_dep} the amplitude ratio and charm strong-phase difference become $r_D^{K3\pi}$ and $\delta_D^{K3\pi}$, respectively, which are quantities averaged over phase space, and the interference terms~\cite{Atwood:2003mj} are multiplied by a coherence factor $\kappa_D^{K3\pi}$.  These parameters have been measured in studies of charm mixing and quantum-correlated $D$-meson decays: $r_D^{K3\pi} = 0.0549 \pm 0.006$, $\delta_D^{K3\pi}=(128\,^{+28}_{-17})^\circ$ and $\kappa_D^{K3\pi}=0.43\,^{+0.17}_{-0.13}$~\cite{LHCb-PAPER-2015-057,Evans:2016tlp}.

In the favoured ADS modes the asymmetries 
\begin{equation}
    \mathcal{A}^{K\pi(\pi\pi)}_{\rm ADS} = \frac{\Gamma(\Bzb\to D(\Km\pip(\pip\pim))\Kstarzb) - \Gamma(\Bz \to D(\Kp\pim(\pip\pim))\Kstarz)}{\Gamma(\Bzb\to D(\Km\pip(\pip\pim))\Kstarzb) + \Gamma(\Bz \to D(\Kp\pim(\pip\pim))\Kstarz)} 
\label{eq:A_ADS}
\end{equation} 
are measured. These modes are expected to exhibit much smaller \CP\ asymmetries than the suppressed decay channels.  

Observables associated with the decay $\Bsb \to D \Kstarz$ are also measured. This decay is expected to exhibit negligible \CP\ violation, but serves as a useful control mode. In this case, for the ADS selection the final state with opposite-sign kaons constitutes the favoured mode, and so the analogously defined asymmetries $\mathcal{A}_{s, \rm ADS}^{\pi K (\pi\pi)}$ are measured. Signal yields are currently too small to permit a study of the suppressed mode.  Finally, the GLW asymmetries $\mathcal{A}^{KK}_{s,\CP}$, $\mathcal{A}^{\pi\pi}_{s,\CP}$ and $\mathcal{A}^{4\pi}_{s,\CP}$, defined analogously to Eq.~\ref{eq:A_hh}, and the ratios $\mathcal{R}^{KK}_{s,\CP}$, $\mathcal{R}^{\pi\pi}_{s,\CP}$ and $\mathcal{R}^{4\pi}_{s,\CP}$, defined analogously to Eq.~\ref{eq:rhhcp}, are determined.

\section{Detector, online selection and simulation}
\label{sec:Detector}
The \lhcb detector~\cite{LHCb-DP-2008-001,LHCb-DP-2014-002} is a single-arm forward
spectrometer covering the \mbox{pseudorapidity} range $2<\eta <5$,
designed for the study of particles containing \bquark or \cquark
quarks. The detector includes a high-precision tracking system
consisting of a silicon-strip vertex detector surrounding the $pp$
interaction region, a large-area silicon-strip detector located
upstream of a dipole magnet with a bending power of about
$4{\mathrm{\,Tm}}$, and three stations of silicon-strip detectors and straw
drift tubes placed downstream of the magnet.
The tracking system provides a measurement of the momentum, \ptot, of charged particles with
a relative uncertainty that varies from 0.5\% at low momentum to 1.0\% at 200\gevc.
The minimum distance of a track to a primary vertex (PV), the impact parameter (IP), 
is measured with a resolution of $(15+29/\pt)\mum$,
where \pt is the component of the momentum transverse to the beam, in\,\gevc.
Different types of charged hadrons are distinguished using information
from two ring-imaging Cherenkov (RICH) detectors. 
Photons, electrons and hadrons are identified by a calorimeter system consisting of
scintillating-pad and preshower detectors, an electromagnetic
and a hadronic calorimeter. Muons are identified by a
system composed of alternating layers of iron and multiwire
proportional chambers.

The online event selection is performed by a trigger, 
which consists of a hardware stage, based on information from the calorimeter and muon
systems, followed by a software stage, which applies a full event
reconstruction. The events considered in the analysis must be triggered at the hardware level when either one of the final-state tracks of the signal decay deposits enough energy in the calorimeter system, or when one of the other tracks in the event, not reconstructed as part of the signal candidate, fulfils any trigger requirement. 
At the software stage, it is required that at least one particle should have high \pt\ and high \chisqip, where \chisqip\ is defined as the difference in the PV fit \chisq\ with and without the inclusion of that particle. A multivariate algorithm~\cite{BBDT} is used to identify secondary vertices consistent with being a two-, three- or four-track $b$-hadron decay. The PVs are fitted with and without the $B$ candidate tracks, and the PV that gives the smallest \chisqip\ is associated with the $B$ candidate.

Simulated events are used to describe the signal mass shapes and compute efficiencies. In the simulation, $pp$ collisions are generated using
\pythia~\cite{Sjostrand:2007gs} with a specific \lhcb\ configuration~\cite{LHCb-PROC-2010-056}. Decays of hadronic particles
are described by \evtgen~\cite{Lange:2001uf}, in which final-state
radiation is generated using \photos~\cite{Golonka:2005pn}. The
interaction of the generated particles with the detector, and its response,
are implemented using the \geant
toolkit~\cite{Allison:2006ve, *Agostinelli:2002hh} as described in
Ref.~\cite{LHCb-PROC-2011-006}.

\section{Offline selection}
\label{sec:selection}

Signal $B$-meson candidates are obtained by combining $D$ and $\Kstarz$ candidates, and are required to have a $\pt$ greater than $5\gevc$, a lifetime greater than $0.2\ps$, and a good-quality vertex fit. The $D$ candidate is reconstructed from the seven different decay modes of interest within a $\pm 25$\mevcc\ window around the known $\Dz$ mass\cite{PDG2018}, and must have a $\pt$ greater than $1.8\gevc$. The $\Kstarz$ candidate is reconstructed from the final state $\Kp\pim$, selected within a $\pm 50$\mevcc\ window around the known $\Kstarz$ mass and with a total $\pt$ of at least $1\gevc$. This mass window is approximately the width of the $\Kstar(892)^0$ resonance\cite{PDG2018}. The helicity angle $\theta^*$, defined as the angle between the $\Kp$ momentum in the $\Kstarz$ rest frame and the $\Kstarz$ momentum in the $\Bz$ rest frame, is required to satisfy $\lvert\cos(\theta^*)\rvert > 0.4$. This requirement removes 60\% of the combinatorial background with a fake $\Kstarz$, while retaining 93\% of the signal. The $D$ and $\Kstarz$ candidates are both required to have a good-quality vertex fit, a significant separation from the PV, and a distance-of-closest-approach between their decay products of less than $0.5\mm$. All charged final-state particles are required to have a good-quality track fit, $p$ greater than $1\gevc$, and $\pt$ greater than $100\mevc$. The $B$ decay chain is refitted~\cite{Hulsbergen:2005pu} with the $D$ mass fixed to its known value and the $B$ meson constrained to originate from its associated PV\@.

Gradient Boosted Decision Trees (BDTs)\cite{Breiman} are used to separate signal from combinatorial background. A shared BDT is employed for the favoured and suppressed two-body ADS modes, and similarly for the four-body ADS modes. Three independent BDTs are used to select the $\Kp\Km$, $\pip\pim$ and $\pip\pim\pip\pim$ decays. All BDTs are trained with samples of simulated $\Bz \to D\Kstarz$ decays as signal and with candidates from the upper $B$ mass sideband ($5800 < m(B) < 6000\mevcc$) in data as background. The discriminating variables in the BDT comprise: the $B$ vertex-fit \chisq; the \chisqip\ of the $B$ and $D$ candidates; the $\chisqip$ and $\pt$ of the $\Kstarz$ products; the angle between the $B$ momentum vector and the vector between the PV and the $B$ decay vertex; and the $\pt$ asymmetry between the $B$ candidate and other tracks from the PV in a cone around the $B$ candidate. 
The \pt\ asymmetry is defined as $(\pt^B - \pt^\mathrm{cone})/(\pt^B + \pt^\mathrm{cone})$, where $\pt^B$ is the transverse momentum of the $B$ candidate and $\pt^\mathrm{cone}$ is the scalar sum of the transverse momenta of all other tracks in the cone.
The radius of the cone is chosen to be 1.5 units in the plane of pseudorapidity and azimuthal angle (expressed in radians). The \pt\ asymmetry is a measure of the isolation of the $B$ candidate. The BDTs applied to $B$ candidates with two-body $D$-meson decays also use the $\pt$ and \chisqip\ of the $D$ decay products. These variables are not included in the BDTs applied to $B$ candidates with four-body $D$ decays to avoid significant changes to the phase-space distribution.

Particle-identification (PID) information from the RICH detectors is used to improve the purity of the different $D$-meson samples. Criteria are chosen such that no candidate can appear in more than one $D$ decay category. A stringent PID requirement is applied to the kaon from the $\Kstarz$ candidate to suppress contamination from $\Bz \to D\pip\pim$ decays, with a pion misidentified as a kaon.

It is possible for both the kaon and pion (or one of the two pions, in the four-body case) from the $D$-meson decay in the favoured mode to be misidentified, and thus pollute the suppressed sample. To eliminate this source of contamination, the $D$ invariant mass is reconstructed with the opposite mass hypothesis for the kaon and pion. Candidates within $\pm 15$\mevcc\ of the known $\Dz$ mass in this alternative reconstruction are vetoed. After this veto, a contamination rate of $\mathcal{O}(0.1\%)$ is expected in the suppressed mode. No veto is applied to remove $\Bz \to D\Kstarz$ decays where both the kaon and pion from the $\Kstarz$ candidate are misidentified, as this background is sufficiently suppressed by the PID requirement on the kaon, leaving a contamination rate of $\mathcal{O}(0.7\%)$ in the suppressed mode.

Additional background can arise from $B^0_{(s)} \to \Dm h^+$ ($h=K, \pi$), $\Dsm \Kp$ or $\Ds \pim$ decays, with $\D_{(s)}^\pm$ decaying to a three-body combination of kaons and pions.
This contamination is removed by imposing a $\pm 15$\mevcc\ veto around the known $D_{(s)}^\pm$ mass in the invariant mass of the relevant three tracks. These vetoes are over 99\% efficient at retaining signal candidates.

A background from charmless $B$ decays that peaks at the same invariant mass as the signal is suppressed by requiring that the flight distance of the $D$ candidate divided by its uncertainty be greater than 3. A further background from $\Bp \to D\Kp$ decays that are mistakenly combined with a random pion from elsewhere in the event contaminates the region in invariant mass above the signal. This background is removed with a veto of $\pm 25$\mevcc\ around the known $\Bp$ mass in the invariant mass of the $D$ meson and the kaon from the $\Kstarz$ candidate. 

\section{Invariant-mass fit}
\label{sec:mass_fit}
The selected data set comprises two LHC runs and seven $D$-meson decay modes. The sample is further divided into $\Bz$ and $\Bzb$ candidates, based on the charge of the kaon from the $\Kstarz$ meson. This gives a total of 26 categories, as the $\pip\pim\pip\pim$ channel is not selected in the Run 1 data. The invariant-mass distributions  are fit simultaneously in these categories with an unbinned extended maximum-likelihood fit. A fit model is developed comprising several signal and background components, which unless otherwise stated are modelled using simulated signal and background samples reconstructed as the signal decay and passing the selection requirements. The components are:

\begin{enumerate}
    \item Signal $\Bz \to D\Kstarz$ and $\Bsb \to D\Kstarz$ decays, described by Cruijff functions~\cite{cruijff} with free means and widths, and tail parameters fixed from simulation.
    \item Combinatorial background, described by an exponential function with a free slope. As the shape is completely free in the fit, no simulation is used to model this background.
    \item Partially reconstructed background from $\Bz \to \Dstar\Kstarz$ and $\Bsb \to \Dstar\Kstarz$ decays, where $\Dstar$ represents either a $\Dstarz$ or $\Dstarzb$ meson. The $\Dstar$ meson decays via $\Dstar \to D \piz$ or $\Dstar \to D \gamma$, where the neutral pion or photon is not reconstructed. These components are described by analytic probability distribution functions constructed from a parabolic function to describe the decay kinematics that is convolved with the sum of two Gaussians with a common mean to describe the detector resolution, as further described in Ref.~\cite{LHCb-PAPER-2017-021}. All shape parameters are fixed from simulation. The form of the parabola depends on both the missed particle and the helicity of the $\Dstar$ meson, which can be equal to zero (longitudinal polarisation) or $\pm 1$ (transverse polarisation).
    \item Partially reconstructed background from $\Bp \to D\Kp\pim\pip$ decays, where the $\pip$ meson is not reconstructed. This background is described by the sum of two Gaussian functions with separate means and a parabola convolved with the sum of two Gaussians with a common mean. All shape parameters are fixed from simulation.
    \item A background from $\Bz \to D\pip\pim$ decays, with one of the pions misidentified as a kaon. This background is described by the sum of two Crystal Ball functions\cite{Skwarnicki:1986xj} with all shape parameters fixed from simulation.
\end{enumerate}

The signal and combinatorial background yields are free parameters for each LHC run. Preliminary studies showed that the ratios of the yields of the partially reconstructed backgrounds with respect to the signal yields are compatible within uncertainties between Runs 1 and 2, and so a single value is used in the fit. The same assumption cannot be made for the misidentified $\Bz \to D\pip\pim$ background, as the yield of this background is affected by the $\pi \to K$ misidentification rate, which can vary between running periods. The proportion of this background with respect to the signal is therefore corrected in Run 2 with respect to Run 1. Studies of simulated signal and background samples determine this correction factor to be $0.928 \pm 0.014$.

The $\Bz \to D\pip\pim$ background is assumed to have no \CP\ asymmetry, as the candidates cannot be tagged as coming from a $\Bz$ or $\Bzb$ decay, and the difference between the $\pip$ and $\pim$ misidentification rates is found to be negligible in simulated samples. Misidentified $\Bz \to D\pip\pim$ decays should therefore contaminate $\Bz$ and $\Bzb$ equally. The $\Bsb \to \Dstar\Kstarz$ background is not expected to exhibit \CP\ violation, so the \CP asymmetry is fixed to zero in the GLW modes but is free in the ADS modes. The yields of the $\Bz \to D\pip\pim$ and $\Bsb \to \Dstar\Kstarz$ backgrounds are free parameters in the ADS modes and fixed in the GLW modes relative to the ADS yields from knowledge of the  $\Dz$ branching fractions\cite{PDG2018} and relative selection efficiencies determined from simulation. The $\Bz \to \Dstar\Kstarz$ background may exhibit \CP\ violation, so the yields of each $D$ decay channel are free parameters, thus allowing for a nonzero \CP\ asymmetry. The relative yields and asymmetries of the $\Bp \to D\Kp\pim\pip$ background are fixed using measurements from Ref.~\cite{LHCb-PAPER-2015-020}.

For both the $\Bz \to \Dstar\Kstarz$ and $\Bsb \to \Dstar\Kstarz$ backgrounds, the relative proportion of partially reconstructed $\Dstar \to D \gamma$ and $\Dstar \to D \pi^0$ decays is fixed by known \Dstarz branching fractions\cite{PDG2018} and relative selection efficiency as determined from simulation. The fraction of longitudinal polarisation is unknown and is therefore a free parameter in the fit. 

\begin{figure}
    \centering
    \setlength{\tabcolsep}{0pt}
    \begin{tabular}{cc}
        \includegraphics[width=0.49\textwidth]{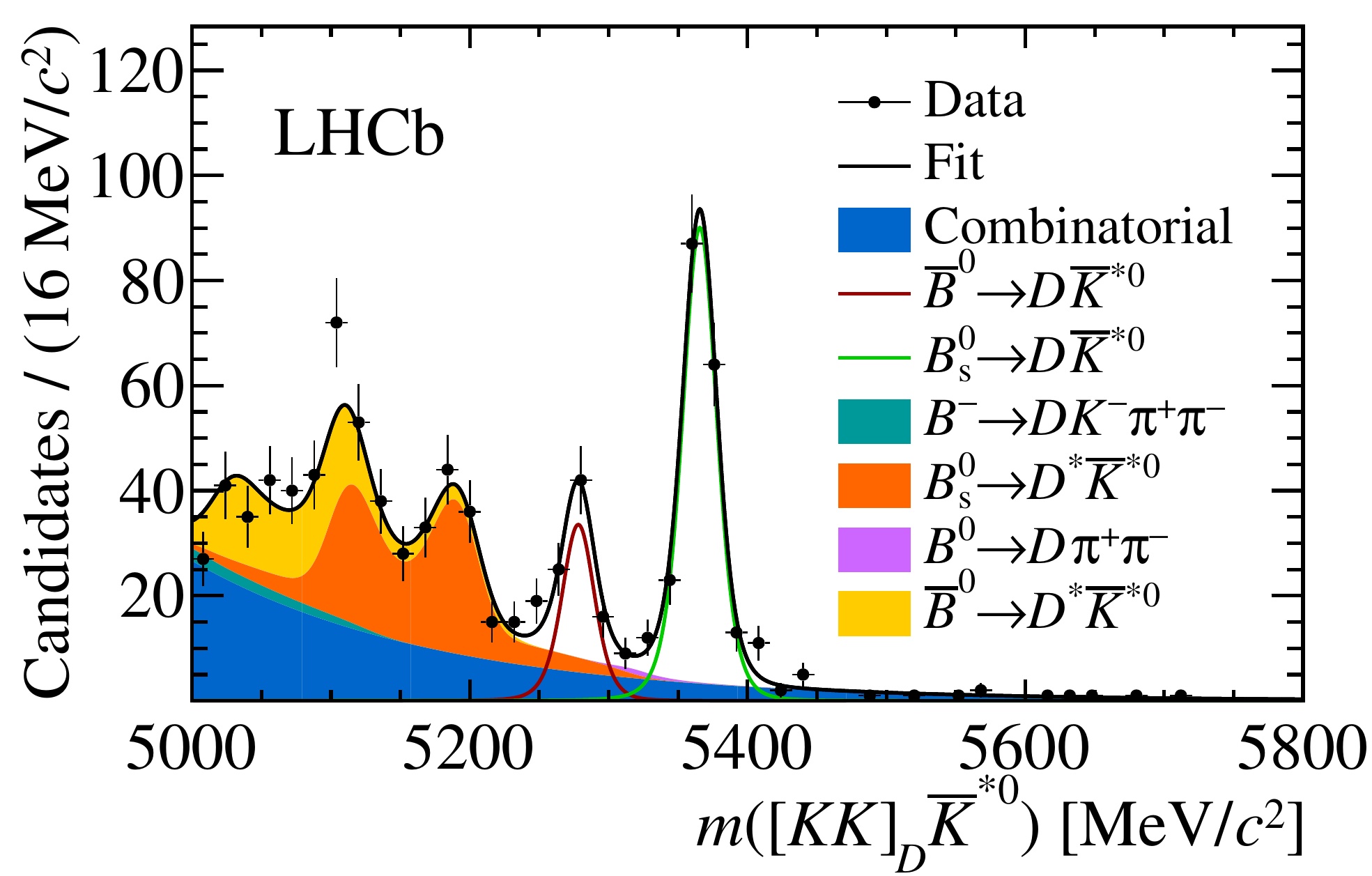} &
        \includegraphics[width=0.49\textwidth]{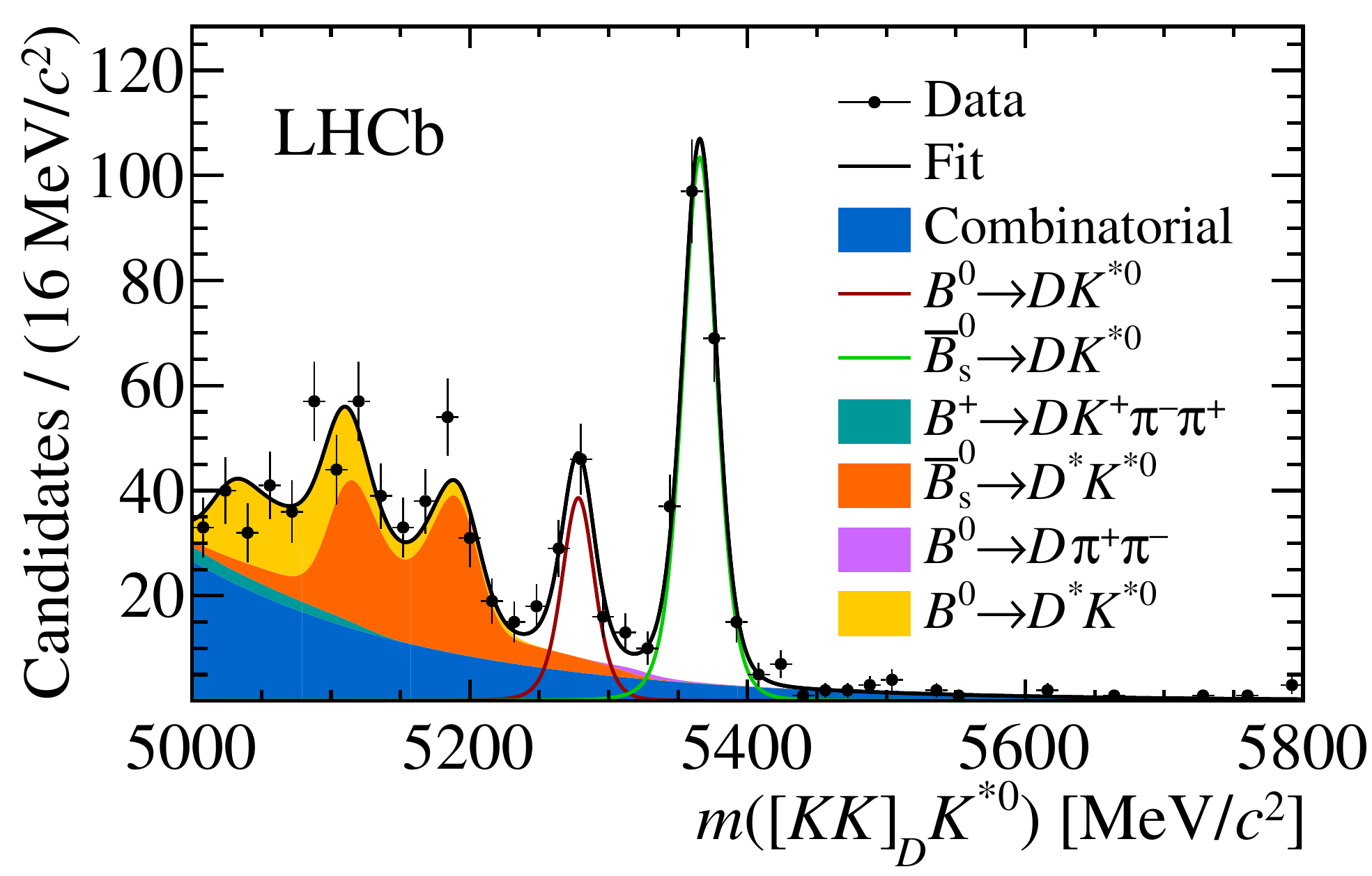} \\
        \includegraphics[width=0.49\textwidth]{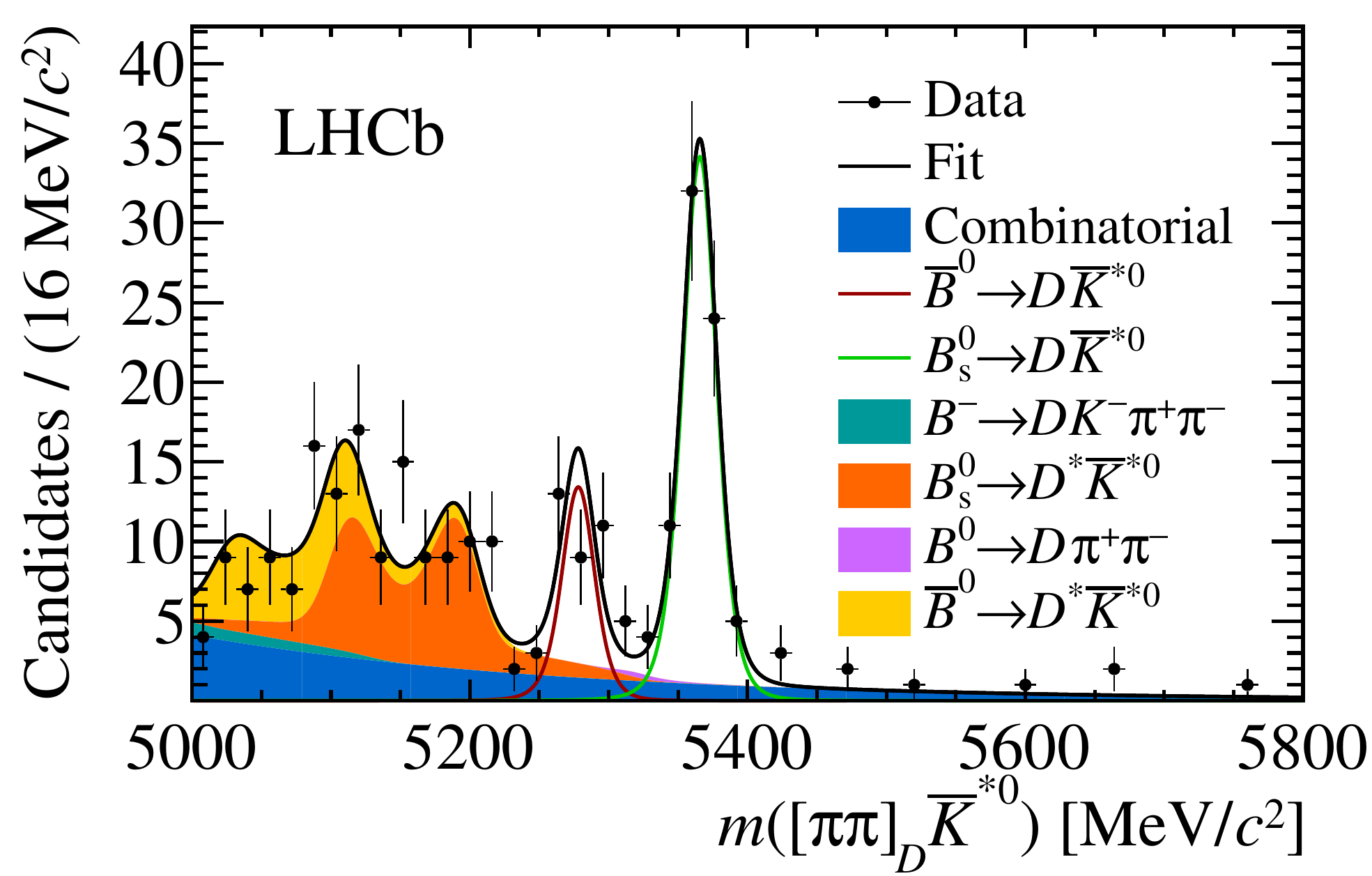} &
        \includegraphics[width=0.49\textwidth]{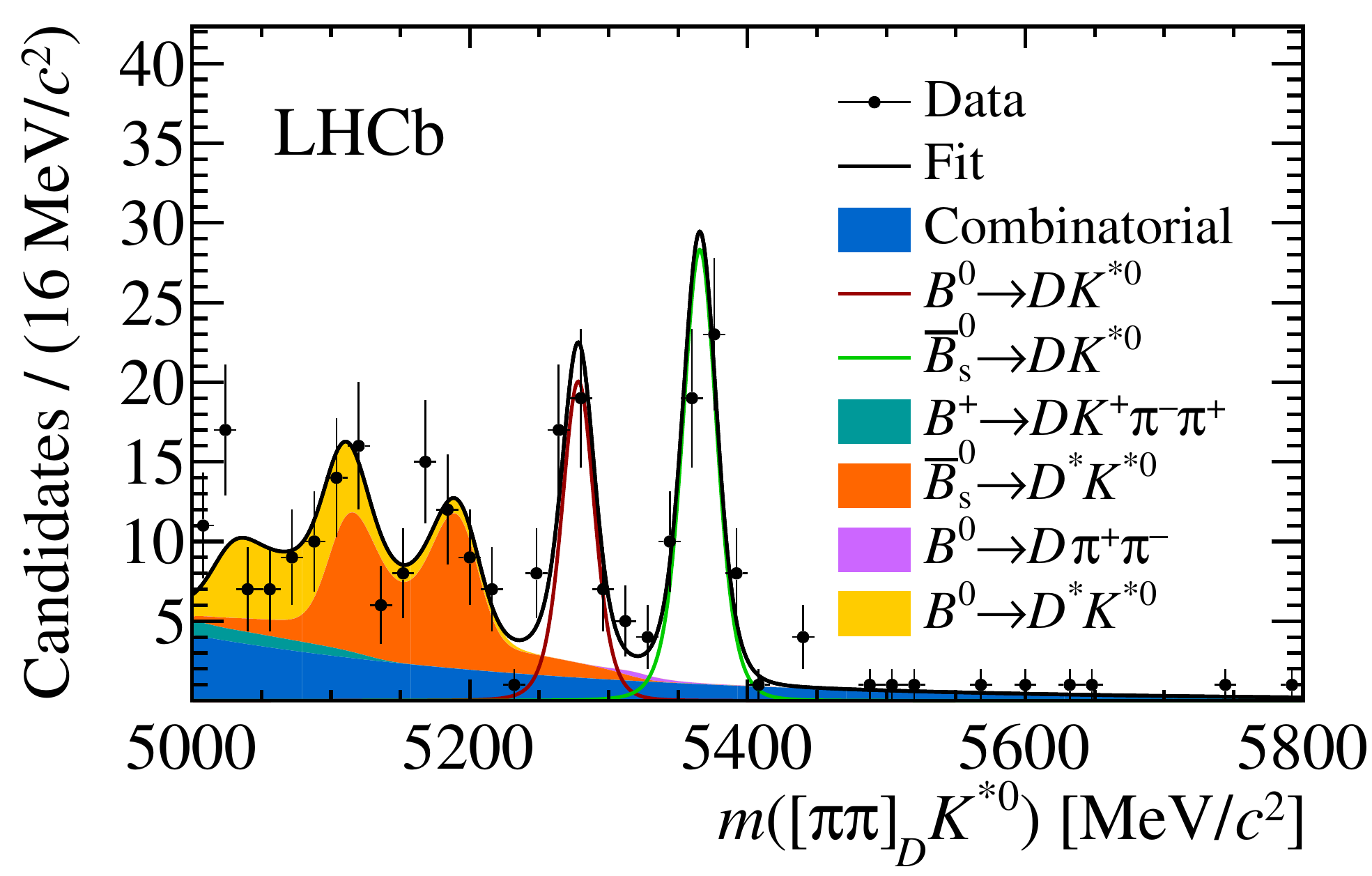} \\
    \end{tabular}
    \caption{
        Invariant-mass distributions (data points with error bars) and results of the fit (lines and coloured areas) for the two-body GLW modes (top left) $\Bzb \to D(\Kp \Km)\Kstarzb$, (top right) $\Bz \to D(\Kp \Km)\Kstarz$, (bottom left) $\Bzb \to D(\pip \pim)\Kstarzb$ and (bottom right) $\Bz \to D(\pip \pim)\Kstarz$. } 
\label{fig:mass_fit_glw_2body}
\end{figure}

\begin{figure}
    \centering
    \setlength{\tabcolsep}{0pt}
    \begin{tabular}{cc}
        \includegraphics[width=0.49\textwidth]{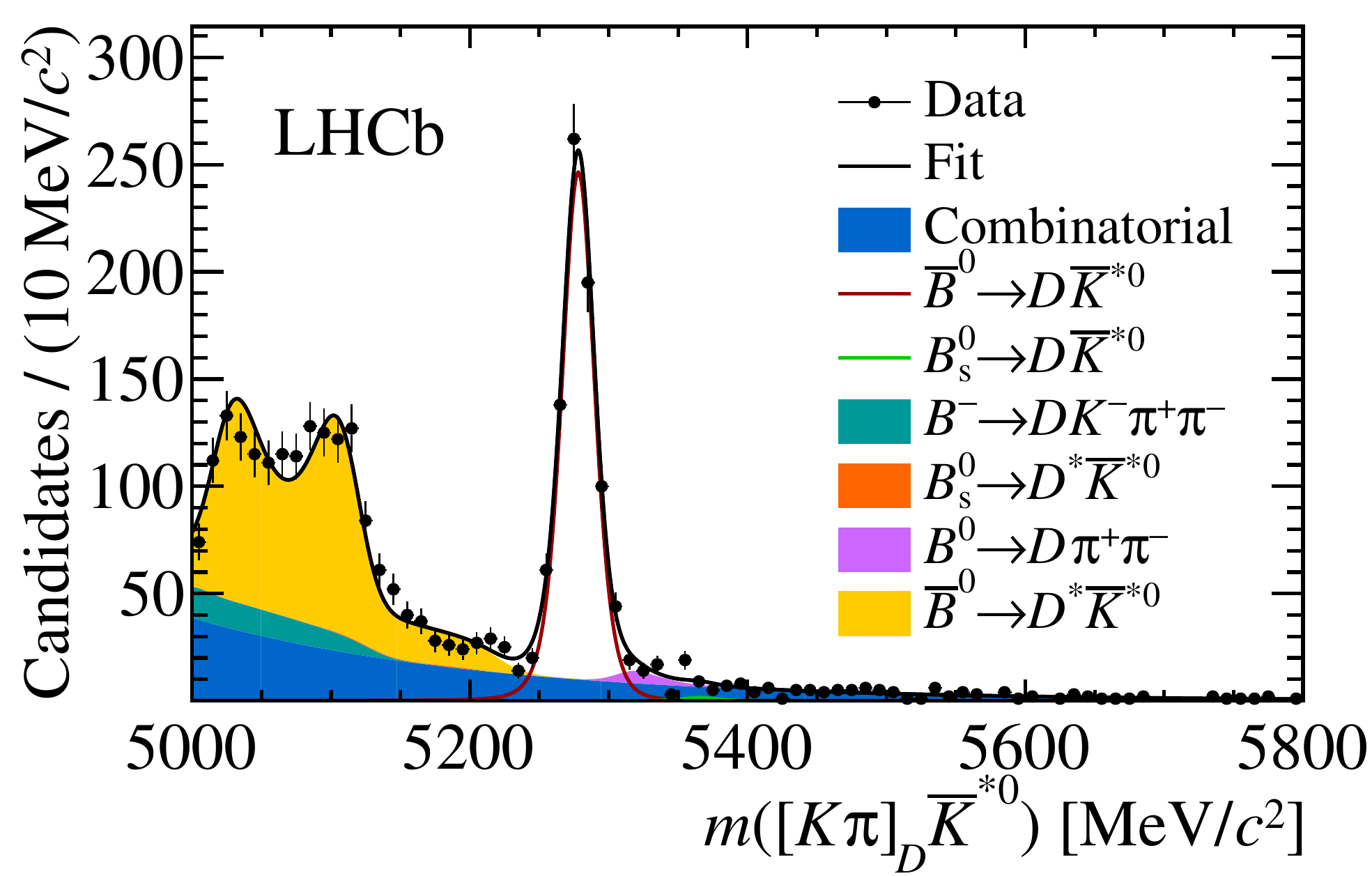} &
        \includegraphics[width=0.49\textwidth]{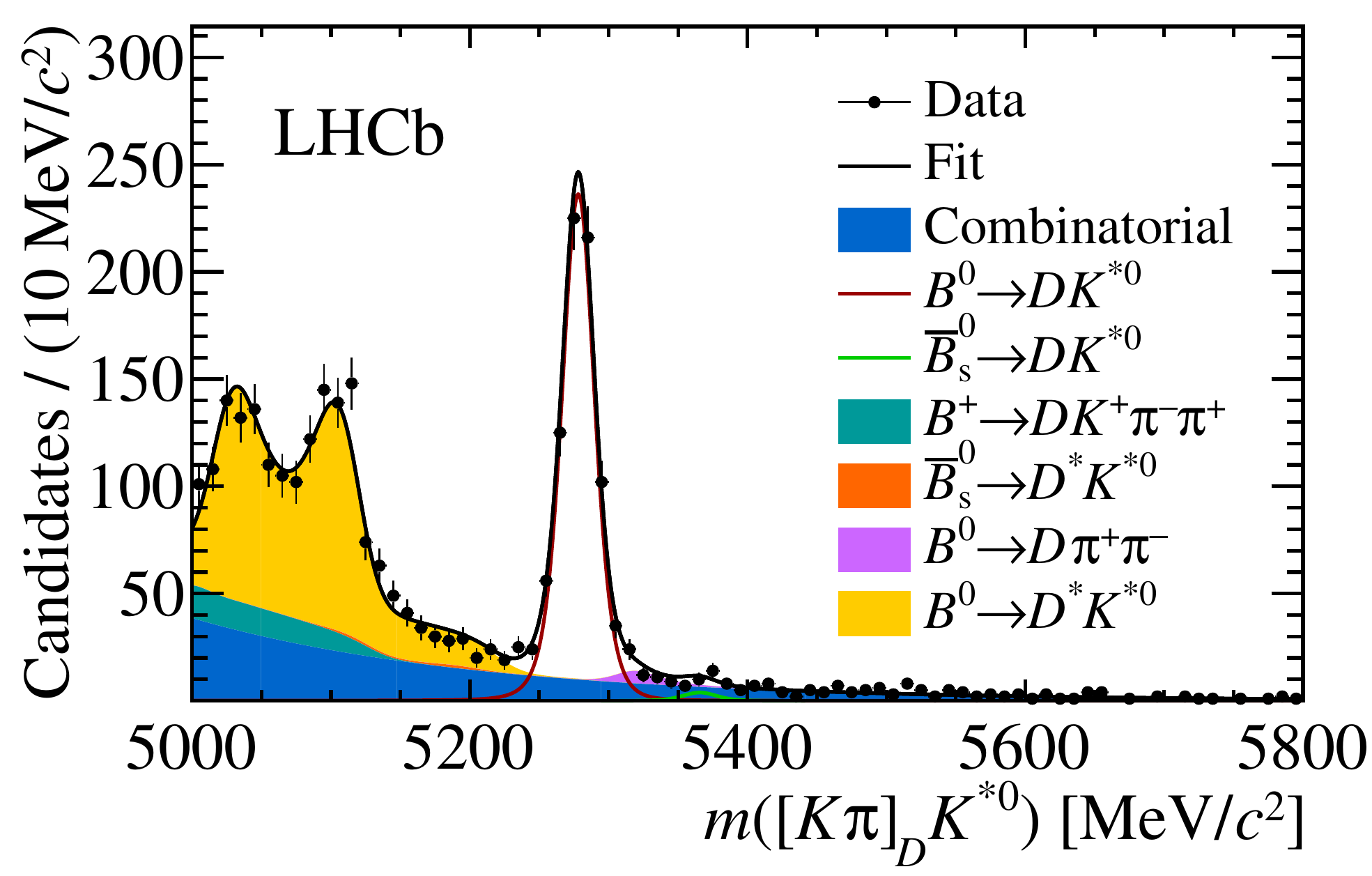} \\
        \includegraphics[width=0.49\textwidth]{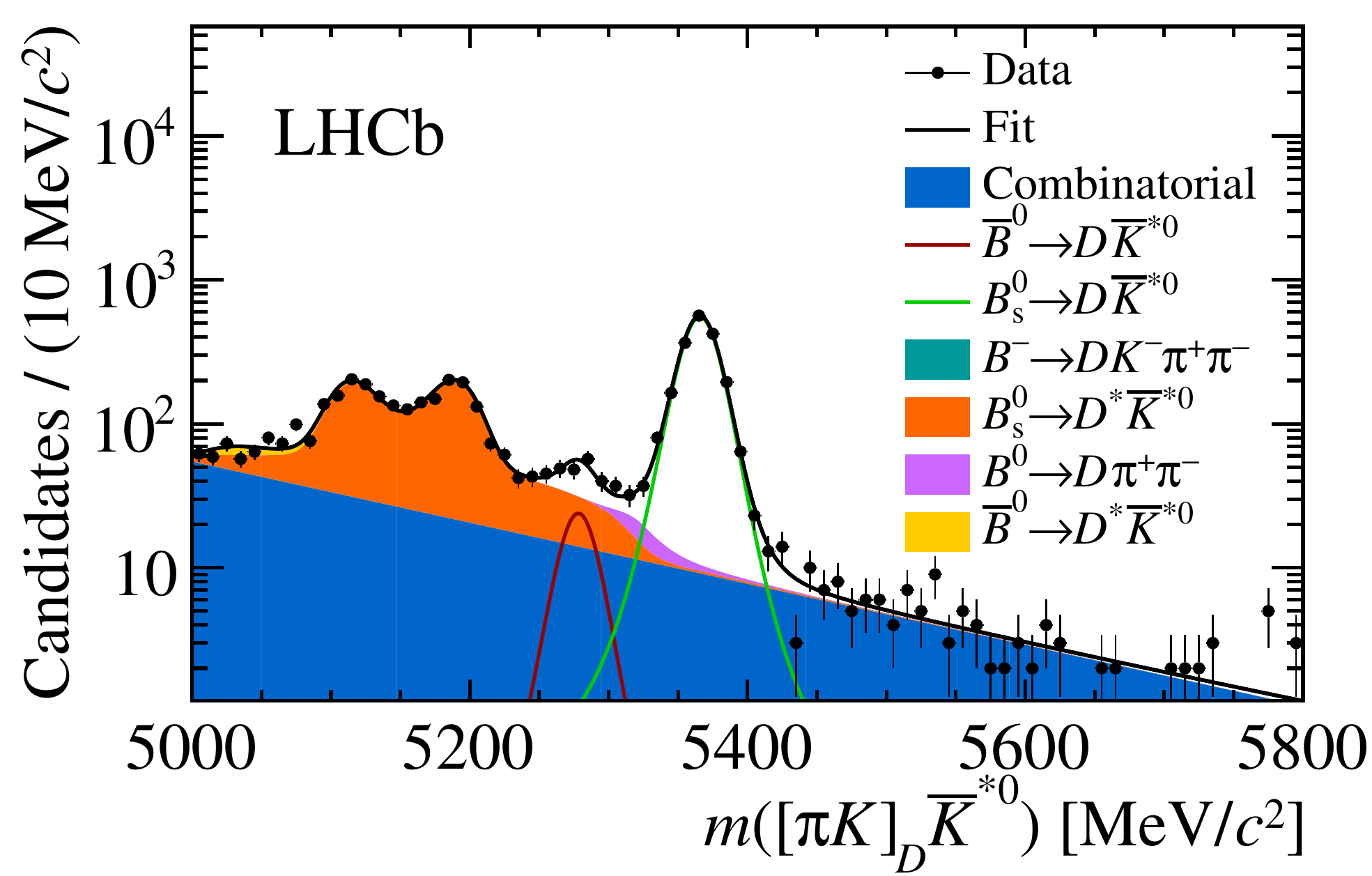} &
        \includegraphics[width=0.49\textwidth]{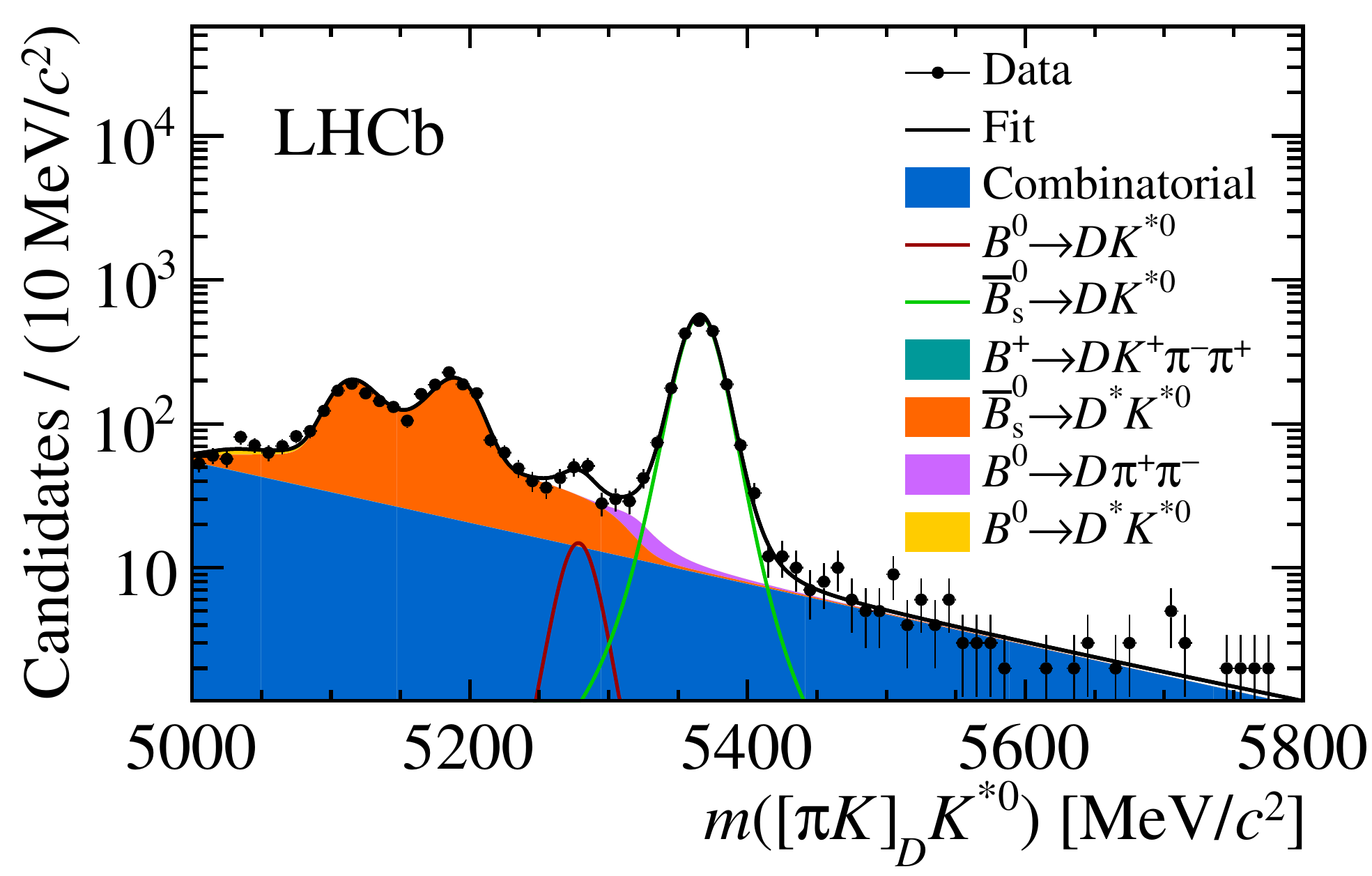} \\
    \end{tabular}
    \caption{
        Invariant-mass distributions (data points with error bars) and results of the fit (lines and coloured areas) for the two-body ADS modes (top left) $\Bzb \to D(\Km \pip)\Kstarzb$, (top right) $\Bz \to D(\Kp \pim)\Kstarz$, (bottom left) $\Bzb \to D(\pim \Kp)\Kstarzb$ and (bottom right) $\Bz \to D(\pip \Km)\Kstarz$. The bottom distributions are shown on a logarithmic scale.}
\label{fig:mass_fit_ads_2body}
\end{figure}

\begin{figure}
    \centering
    \setlength{\tabcolsep}{0pt}
    \begin{tabular}{cc}
        \includegraphics[width=0.49\textwidth]{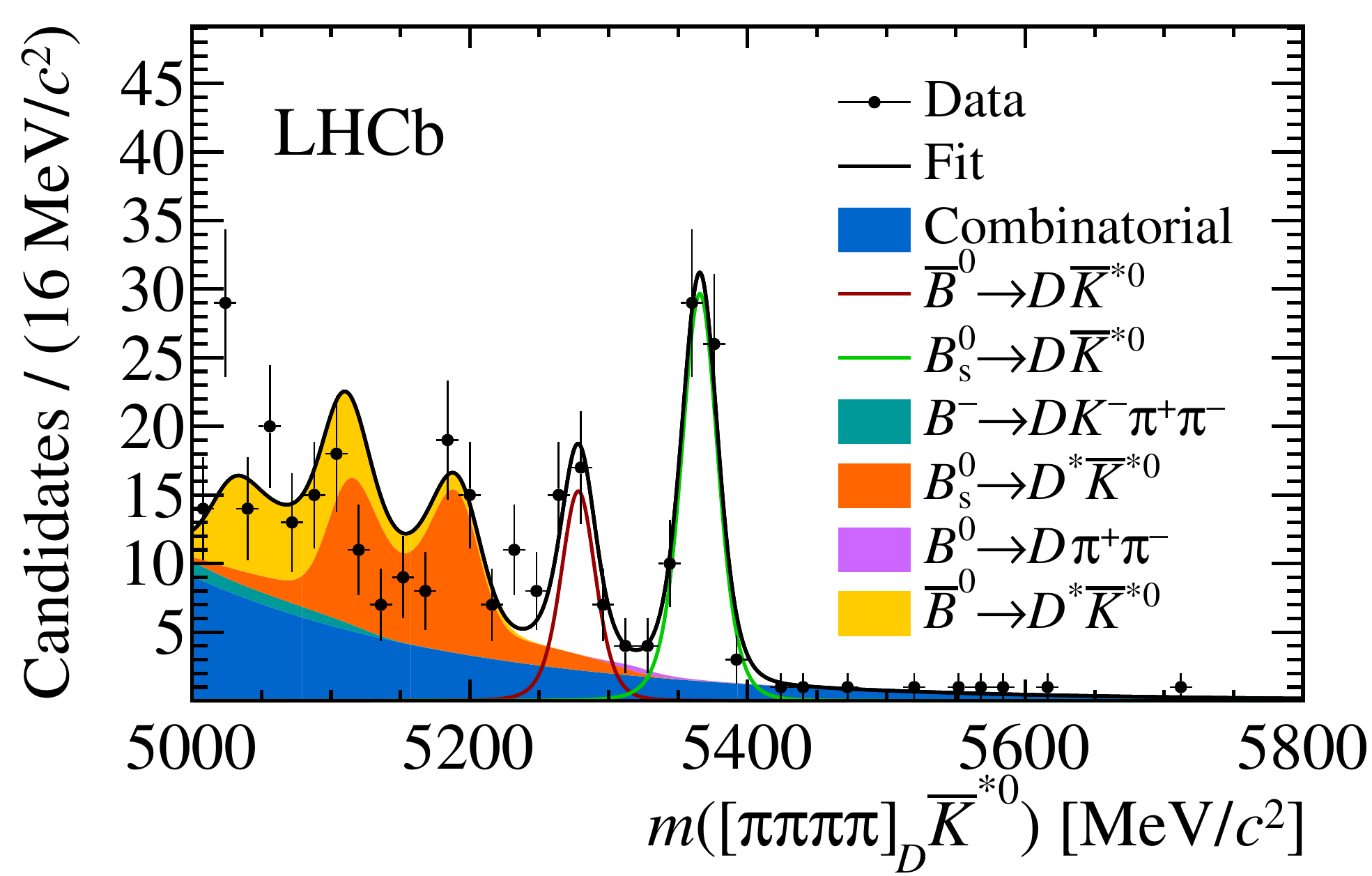} &
        \includegraphics[width=0.49\textwidth]{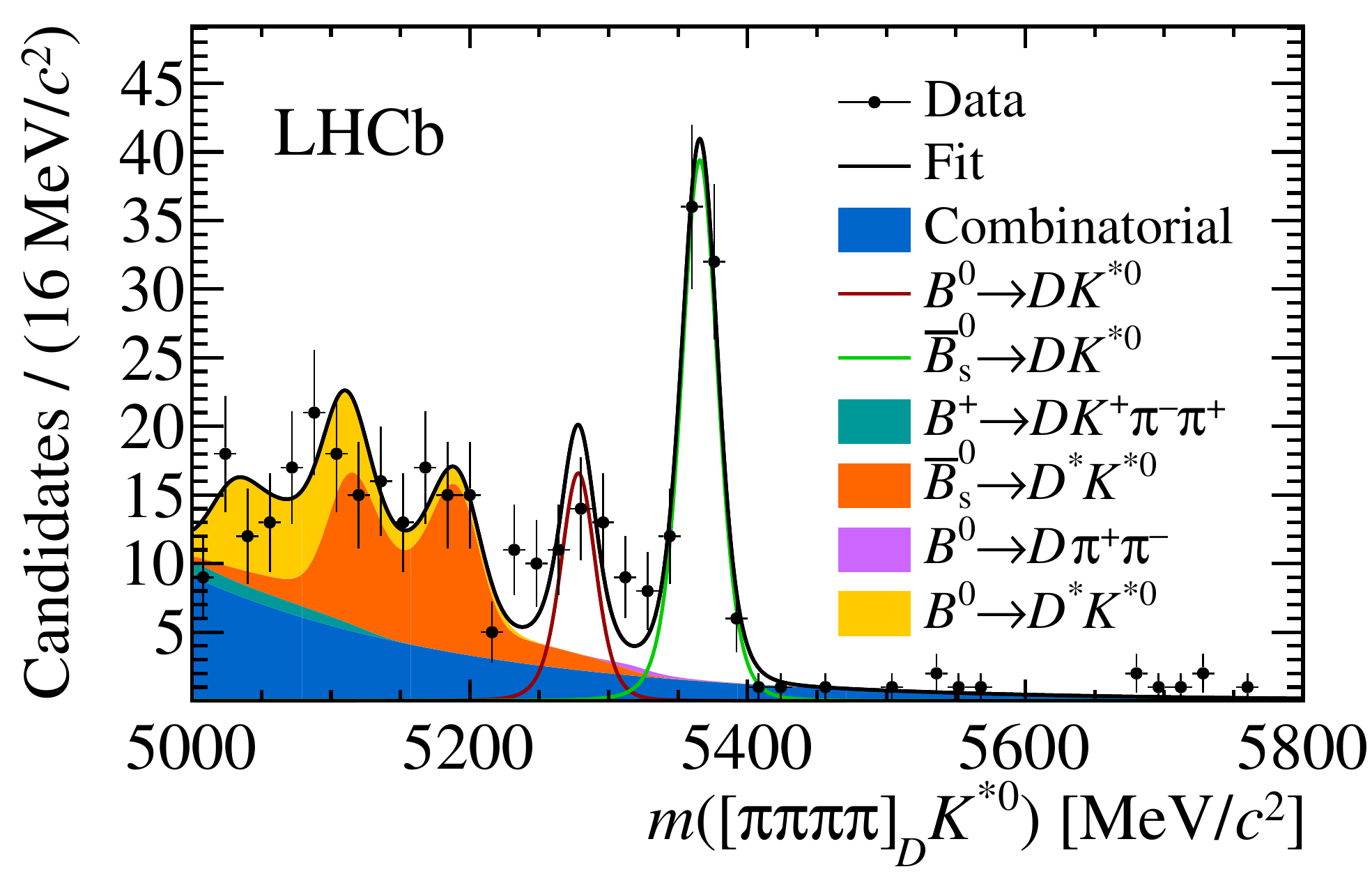} \\
    \end{tabular}
    \caption{
    Invariant-mass distributions (data points with error bars) and results of the fit (lines and coloured areas) for the four-body GLW mode (left) $\Bzb \to D(\pip\pim\pip\pim)\Kstarzb$, (right) $\Bz \to D(\pip\pim\pip\pim)\Kstarz$.} 
\label{fig:mass_fit_glw_4body}
\end{figure}

\begin{figure}
    \centering
    \setlength{\tabcolsep}{0pt}
    \begin{tabular}{cc}
        \includegraphics[width=0.49\textwidth]{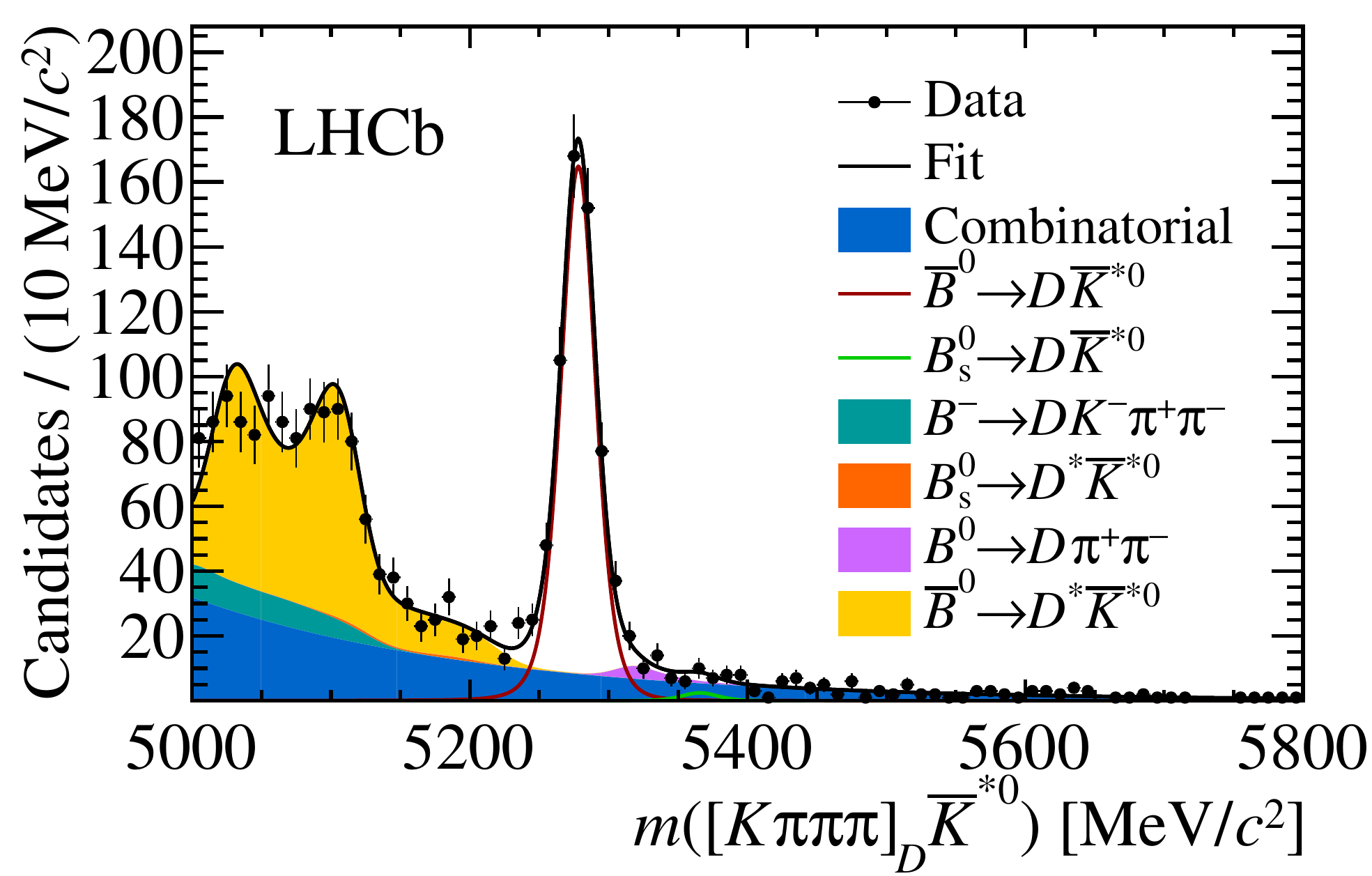} &
        \includegraphics[width=0.49\textwidth]{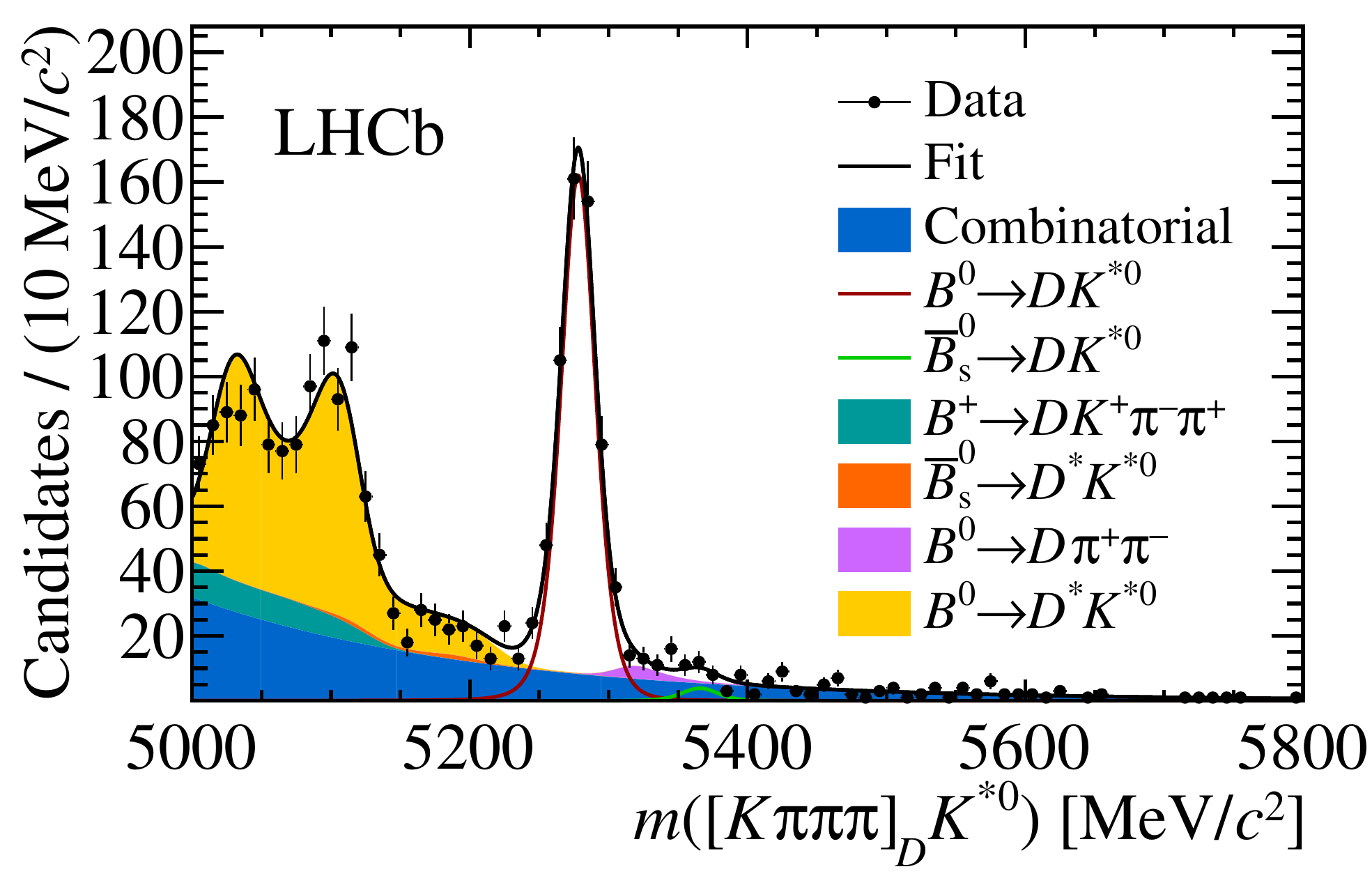} \\
        \includegraphics[width=0.49\textwidth]{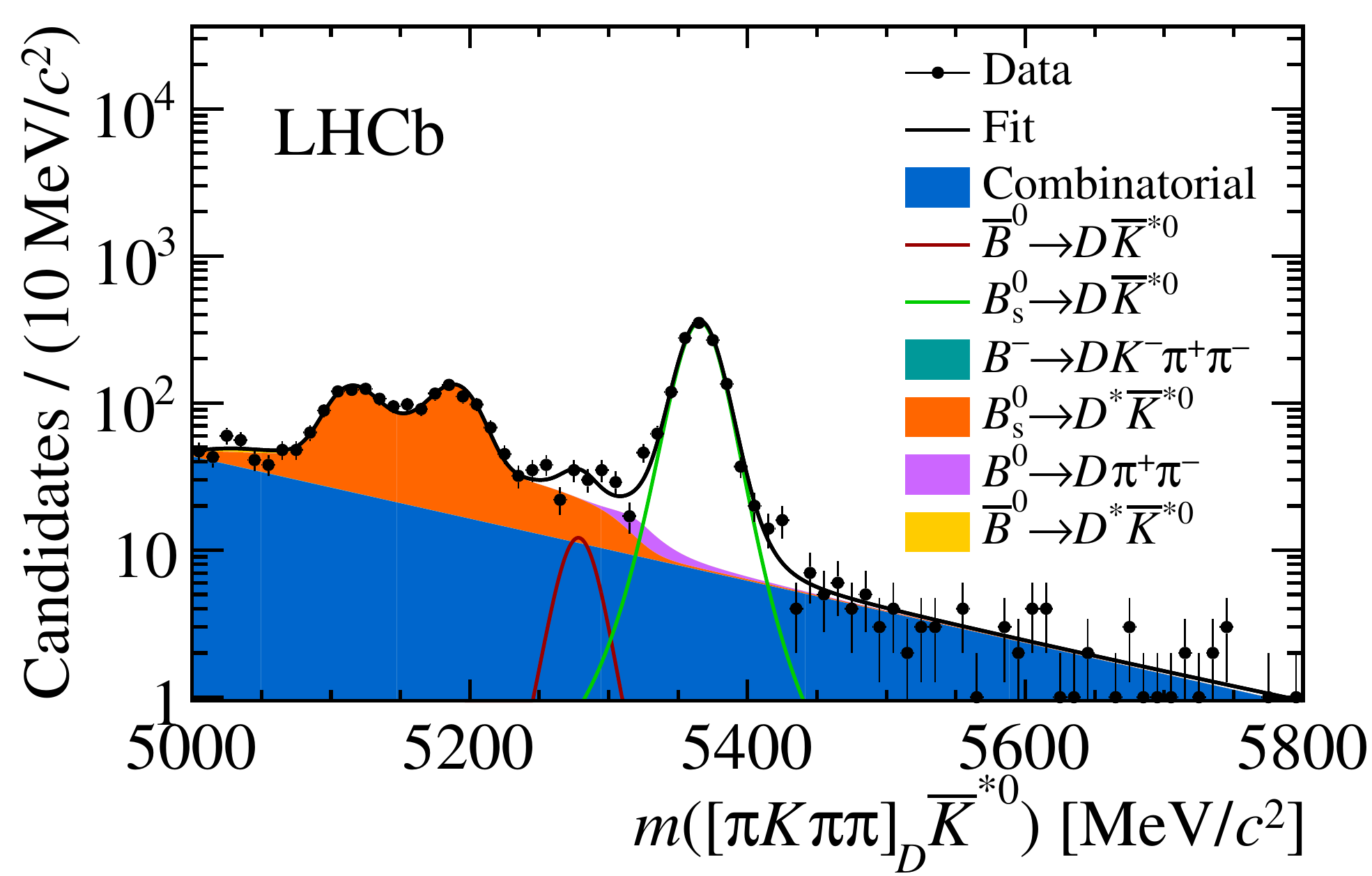} &
        \includegraphics[width=0.49\textwidth]{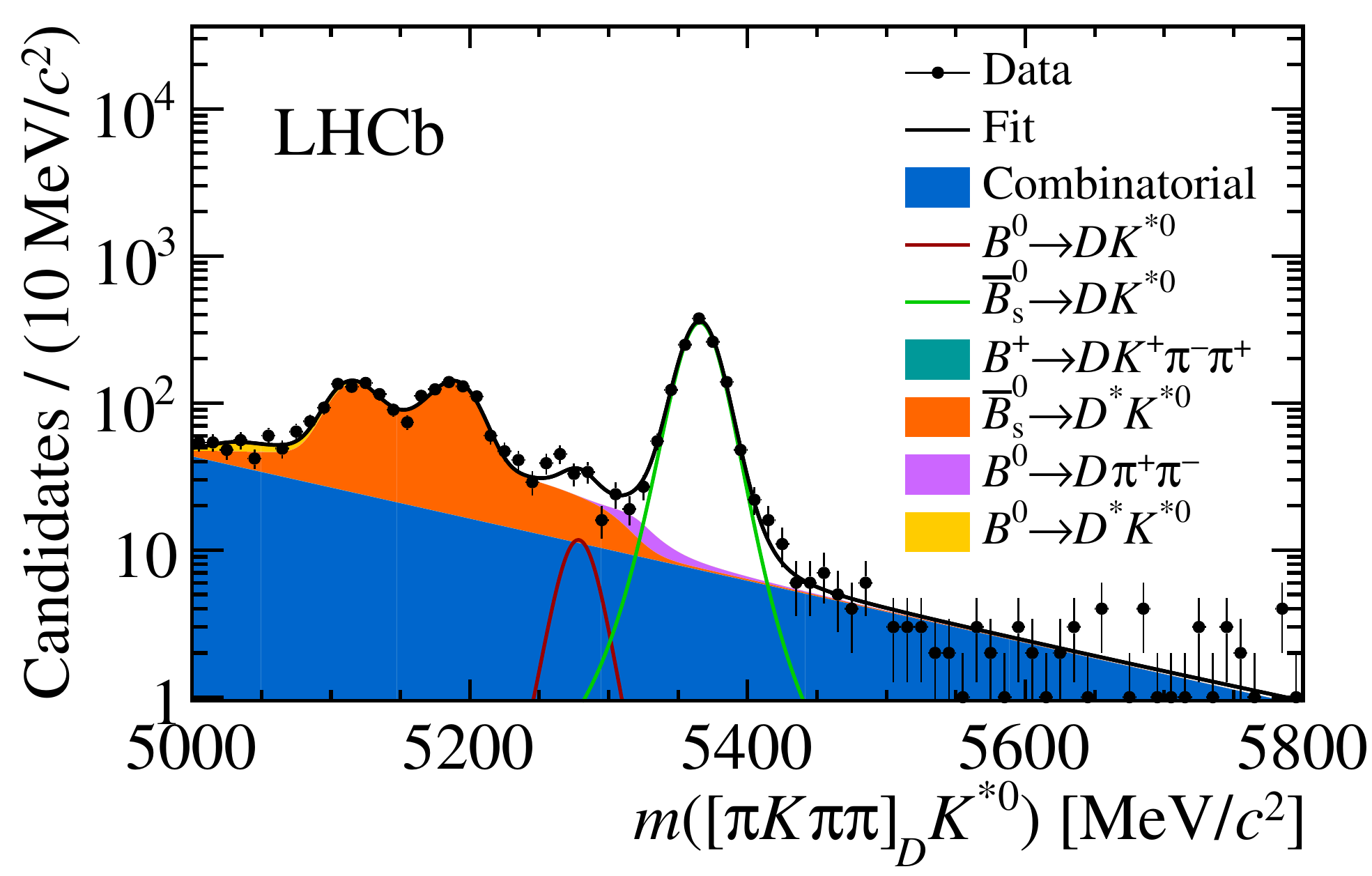} \\
    \end{tabular}
    \caption{
    Invariant-mass distributions (data points with error bars) and results of the fit (lines and coloured areas) for the four-body ADS modes (top left) $\Bzb \to D(\Km \pip \pim \pip)\Kstarzb$, (top right) $\Bz \to D(\Kp \pim \pip \pim)\Kstarz$, (bottom left) $\Bzb \to D(\pim \Kp \pip \pim)\Kstarzb$ and (bottom right) $\Bz \to D(\pip \Km \pim \pip)\Kstarz$. The bottom distributions are shown on a logarithmic scale.}
\label{fig:mass_fit_ads_4body}
\end{figure}

Figures~\ref{fig:mass_fit_glw_2body} to~\ref{fig:mass_fit_ads_4body} show the invariant mass distributions and the fitted shapes for the various components. Table~\ref{tab:yields} gives the signal yields for each $D$ final state. The fit strategy is validated by pseudoexperiments, and is found to be reliable and unbiased for all free parameters.

The observables introduced in Sect.~\ref{sec:observables} are determined directly from the fit. The ratios and asymmetries between the raw yields are corrected for efficiency differences, and production and detection asymmetries. To obtain the ratios $\mathcal{R}^{hh}_{\CP}$ and $\mathcal{R}^{4\pi}_{\CP}$, the raw ratios are normalised using the corresponding $\Dz$ branching fractions. These corrections are discussed further in Sect.~\ref{sec:systematics}.


\begin{table}
    \caption{Summary of signal yields. The uncertainties are statistical.
    } 
    \centering
    \setlength{\tabcolsep}{4pt}
    \begin{tabular}{lr@{}c@{}lr@{}c@{}l}
        \toprule
        Decay channel & \multicolumn{3}{c}{$\Bzb$ yield} & \multicolumn{3}{c}{$\Bz$ yield} \\
        \midrule
        $\Bz \to D(\Kp\Km)\Kstarz$ & $67 $&\,$\pm$\,  &$ 10$ & $77 $& \,$\pm$\,  &$ 11$ \\
        $\Bz \to D(\pip\pim)\Kstarz$ & $27 $&\,$\pm$\,  &$ \phantom{0}6$ & $40 $& \,$\pm$\,  &$ \phantom{0}7$ \\
        $\Bz \to D(\pip\pim\pip\pim)\Kstarz$ & $32 $&\,$\pm$\,  &$ \phantom{0}7$ & $35 $& \,$\pm$\,  &$ \phantom{0}8$ \\
        $\Bz \to D(\Kp\pim)\Kstarz$ & $786 $&\,$\pm$\,  &$ 29$ & $754 $& \,$\pm$\,  &$ 29$ \\
        $\Bz \to D(\pip \Km)\Kstarz$ & $76 $&\,$\pm$\,  &$ 16$ & $47 $& \,$\pm$\,  &$ 15$ \\
         $\Bz \to D(\Kp \pim\pip\pim)\Kstarz$ & $557 $&\,$\pm$\,  &$ 25$ & $548$& \,$\pm$\,  &$ 25$ \\
        $\Bz \to D(\pip \Km \pip \pim)\Kstarz$ & $41 $&\,$\pm$\,  &$ 14$ & $40 $& \,$\pm$\,  &$ 14$ \\       
        \bottomrule
    \end{tabular}
\label{tab:yields}
\end{table}

\section{Correction factors and systematic uncertainties}
\label{sec:systematics}

The measured observables are either asymmetries or ratios of yields between similar final states, and are thus robust against systematic biases.  Nonetheless, small differences in efficiencies between numerator and denominator mean that correction factors must be applied to the ratios, apart from the case of $\mathcal{R}_{\pm}^{\pi  K(\pi\pi)}$ where an identical selection is used for both the suppressed and favoured ADS modes.
The selection efficiencies are computed from simulated signal samples, which are weighted in the transverse momentum and pseudorapidity of the $B$ meson to agree with the data distributions. 
The efficiencies of the PID requirements between different charges and $D$ final states are evaluated using calibration samples, which are weighted to match the momentum and pseudorapidity of the simulated signal samples.
Uncertainties are assigned due to the finite size of the simulated samples, and for possible biases introduced by the binning schemes used in the reweighting of the calibration sample and the background-subtraction procedure used for these samples.

As can be seen from Eq.~\ref{eq:rhhcp}, determining $\mathcal{R}^{hh}_{\CP}$ requires normalising the measured ratio of yields by a ratio of $\Dz$ branching fractions. These branching fractions are taken from Ref.~\cite{PDG2018} and the uncertainties are propagated to the observables.

The raw observables are corrected for detection asymmetry, which is predominantly caused by the shorter interaction length of $\Km$ mesons compared with $\Kp$ mesons. The difference between the kaon and pion detection asymmetries, $A_D(\Km\pip)$, is computed following the method used in Ref.~\cite{LHCb-PAPER-2014-013}. Raw charge asymmetries $A_{\rm raw}(\Km\pip\pip)$ and $A_{\rm raw}(\Kzb\pip)$ are measured for the decays $\Dp \to \Km\pip\pip$ and $\Dp \to \Kzb\pip$, respectively. These asymmetries are determined using calibration samples which are weighted to match the kinematics of the kaons and pions in the signal data set. The value of $A_D(\Km\pip)$ is calculated from $A_D(\Km\pip) = A_{\rm raw}(\Km\pip\pip) - A_{\rm raw}(\Kzb\pip) + A_D(\Kz)$, where $A_D(\Kz)$ is the measured value of the detection asymmetry in the decay $\Kz \to \pip\pim$,  giving $A_D(\Km\pip) = (-0.92 \pm 0.20)$\% in Run 1 and $(-1.0 \pm 0.6)$\% in Run 2.  A correction of $A_D(\Km\pip)$ is applied to the observables for each $\Kpm\pimp$ pair in the final state.  

The observables are also corrected for the asymmetry in the production of $\Bz$ and $\Bzb$ mesons within the acceptance of the analysis, $A_\mathrm{prod}$. This asymmetry has been measured in bins of  $B$-meson momentum and pseudorapidity in Run 1\cite{LHCb-PAPER-2016-062}.  A weighted average based on the kinematical distributions of simulated signal gives $A_\mathrm{prod} = (-0.8 \pm 0.5)$\%. The same central value is applied for Run 2, with the uncertainty doubled in order to account for a possible change in asymmetry due to the higher collision energy.

Uncertainties are assigned to account for the shape parameters that are fixed in the invariant-mass model. The values of these fixed parameters derive from fits to simulated samples, and so the uncertainties on these fits are propagated to the mass model.
The fixed tail parameters of the signal shape are treated as a single source of systematic uncertainty. Uncertainties due to all fixed parameters related to the background shapes are treated simultaneously, apart from those for the partially reconstructed $\Bsb \to \Dstar\Kstarz$ decays, which are an important source of background that overlaps with the signal region, and are therefore treated separately.

Uncertainties are also considered for other fixed parameters in the fit. These are the relative proportion of partially reconstructed $\Dstar \to D \piz$ and $\Dstar \to D \gamma$ decays, the correction to the relative yield of misidentified $\Bz \to D\pip\pim$ decays between Run 1 and Run 2, and the relative yields and \CP\ asymmetries of the partially reconstructed $\Bp \to D\Kp\pim\pip$ background. For the latter, the uncertainties taken from Ref.~\cite{LHCb-PAPER-2015-020} are doubled to account for the fact that there are possible differences in the phase-space acceptance between the two analyses.

A study of the invariant-mass sidebands of the $D$ candidates is performed in order to search for evidence of any residual charmless background which would also contaminate the $B$ signal region.  This sideband study is performed after imposing the flight distance cut on the $D$ candidate, but without applying the BDT selection, as this may not have a uniform acceptance in $D$ mass.  Regions of the sidebands where there are known reflections from $D$-meson decays with misidentified final products are excluded.  No significant signals are found from charmless decays in any mode.   The measured yields are extrapolated into the signal region and taken as the central values from which many pseudoexperiments with an added charmless background component are simulated. These data sets are fitted using the nominal fit model which neglects the new background contribution, and a systematic uncertainty is assigned based on the measured bias.

Table~\ref{tab:systematics} gives the systematic uncertainties for each observable. Systematic uncertainties which are more than two orders of magnitude smaller than the statistical uncertainty are considered to be negligible and ignored. The non-negligible uncertainties are added in quadrature to give the total systematic uncertainty, which in all cases is considerably smaller than the statistical uncertainty.

The acceptance for the four-body $D$-decay modes is not fully uniform across phase space. Studies performed with amplitude models of these decays indicate that, at the current level of sensitivity, a nonuniform acceptance does not lead to any significant bias when the observables are interpreted in terms of $\gamma$ and the other underlying physics parameters. No systematic uncertainty is assigned.

\begin{sidewaystable}
  \small
  \centering
  \caption{Systematic uncertainties for the observables.  Uncertainties are shown if they are larger than 1\% of the statistical uncertainty. The total systematic uncertainty is calculated by summing all sources in quadrature. Statistical uncertainties are given for reference.}
  \begin{tabular}{lcccccccccccc}
      \toprule
      Source & $\mathcal{A}_{\CP}^{KK}$ & $\mathcal{A}_{\CP}^{\pi\pi}$ & $\mathcal{R}_{\CP}^{KK}$ & $\mathcal{R}_{\CP}^{\pi\pi}$ & $\mathcal{A}_{\CP}^{4\pi}$ & $\mathcal{R}_{\CP}^{4\pi}$ & $\mathcal{R}_+^{\pi K}$ & $\mathcal{R}_-^{\pi K}$ & $\mathcal{R}_+^{\pi K\pi\pi}$ & $\mathcal{R}_-^{\pi K\pi\pi}$ & $\mathcal{A}_{\rm ADS}^{K\pi}$ & $\mathcal{A}_{\rm ADS}^{K\pi\pi\pi}$ \\
      \midrule
      Selection efficiency                    & -     & -     & 0.008 & 0.011 & -     & 0.012 & -                 & -                 & -                 & -                 & -     & -                 \\
      PID efficiency                          & 0.002 & -     & 0.004 & 0.004 & 0.002 & 0.007 & -                 & -                 & -                 & -                 & 0.002 & \kern11.5pt 0.003 \\
      Branching ratios                        & -     & -     & 0.017 & 0.025 & -     & 0.031 & -                 & -                 & -                 & -                 & -     & -                 \\
      Production asymmetry                    & 0.006 & 0.006 & -     & -     & 0.010 & -     & -                 & -                 & -                 & -                 & 0.006 & \kern11.5pt 0.006 \\
      Detection asymmetry                     & 0.004 & 0.004 & 0.004 & 0.007 & 0.007 & 0.007 & $<0.001$          & $<0.001$          & $<0.001$          & $<0.001$          & 0.008 & \kern11.5pt 0.008 \\
      Signal shape parameters                 & -     & -     & -     & -     & -     & -     & $<0.001$          & $<0.001$          & $<0.001$          & $<0.001$          & -     & -                 \\
      $B^0_s \to D^* K^{*0}$ shape parameters & -     & -     & 0.001 & -     & -     & -     & $<0.001$          & $<0.001$          & $<0.001$          & $<0.001$          & -     & -                 \\
      Other background shape parameters       & -     & -     & -     & 0.003 & -     & 0.003 & $<0.001$          & \kern11.5pt 0.001 & \kern11.5pt 0.001 & \kern11.5pt 0.002 & -     & -                 \\
      $D^* \to D^0 \gamma/\pi^0$ inputs       & -     & -     & 0.002 & -     & -     & 0.002 & \kern11.5pt 0.002 & \kern11.5pt 0.002 & \kern11.5pt 0.002 & \kern11.5pt 0.002 & -     & -                 \\
      $B\to D\pi\pi$ PID correction           & -     & -     & -     & -     & 0.006 & -     & $<0.001$          & $<0.001$          & -                 & -                 & -     & $<0.001$          \\
      $B\to DK\pi\pi$ inputs                  & -     & -     & 0.001 & 0.002 & -     & 0.002 & -                 & -                 & -                 & -                 & -     & -                 \\
      Charmless background                    & 0.003 & 0.002 & -     & 0.003 & 0.004 & 0.011 & $<0.001$          & $<0.001$          & -                 & $<0.001$          & 0.002 & \kern11.5pt 0.001 \\
      \midrule
      Total systematic                        & 0.008 & 0.008 & 0.020 & 0.029 & 0.014 & 0.037 & \kern11.5pt 0.002 & \kern11.5pt 0.003 & \kern11.5pt 0.002 & \kern11.5pt 0.003 & 0.010 & \kern11.5pt 0.010 \\
      \midrule
      Statistical                             & 0.10  & 0.14  & 0.10  & 0.19  & 0.15  & 0.16  & \kern11.5pt 0.021 & \kern11.5pt 0.021 & \kern11.5pt 0.026 & \kern11.5pt 0.025 & 0.027 & \kern11.5pt 0.032 \\      \bottomrule
  \end{tabular}
\label{tab:systematics}
\end{sidewaystable}

\section{Results and discussion}
\label{sec:results}
The measured values for the principal observables are

\begin{center}
\setlength{\tabcolsep}{2pt}
\begin{tabular}{rclclcl}
$\mathcal{A}_{\CP}^{KK}$ &= & $-0.05$ & $\pm$ & $0.10$ & $\pm$ & $0.01$, \\
$\mathcal{A}_{\CP}^{\pi\pi}$ &= & $-0.18$ & $\pm$ & $0.14$ & $\pm$ & $0.01$, \\
$\mathcal{R}_{\CP}^{KK}$ &= & $\phantom{+}0.92$ & $\pm$ & $0.10$ & $\pm$ & $0.02$, \\
$\mathcal{R}_{\CP}^{\pi\pi}$ &= & $\phantom{+}1.32$ & $\pm$ & $0.19$ & $\pm$ & $0.03$, \\
$\mathcal{A}_{\CP}^{4\pi}$ &= & $-0.03$ & $\pm$ & $0.15$ & $\pm$ & $0.01$, \\
$\mathcal{R}_{\CP}^{4\pi}$ &= & $\phantom{+}1.01$ & $\pm$ & $0.16$ & $\pm$ & $0.04$, \\
$\mathcal{R}_+^{\pi K}$ &= & $\phantom{+}0.064$ & $\pm$ & $0.021$ & $\pm$ & $0.002$, \\
$\mathcal{R}_-^{\pi K}$ &= & $\phantom{+}0.095$ & $\pm$ & $0.021$ & $\pm$ & $0.003$, \\
$\mathcal{R}_+^{\pi K\pi\pi}$ &= & $\phantom{+}0.074$ & $\pm$ & $0.026$ & $\pm$ & $0.002$, \\
$\mathcal{R}_-^{\pi K\pi\pi}$ &= & $\phantom{+}0.072$ & $\pm$ & $0.025$ & $\pm$ & $0.003$, \\
$\mathcal{A}_{\rm ADS}^{K\pi}$ &= & $\phantom{+}0.047$ & $\pm$ & $0.027$ & $\pm$ & $0.010$, \\
$\mathcal{A}_{\rm ADS}^{K\pi\pi\pi}$ &= & $\phantom{+}0.037$ & $\pm$ & $0.032$ & $\pm$ & $0.010$, \\
\end{tabular}
\end{center}

\noindent where the first uncertainty is statistical, and the second systematic.
The values of $\mathcal{R}^{\pi K}_{\pm}$ and $\mathcal{R}^{\pi K \pi\pi}_{\pm}$ are used to calculate the suppressed-mode ADS observables, which are found to be

\begin{center}
\setlength{\tabcolsep}{2pt}
\begin{tabular}{rclclcl}
$\mathcal{A}_\mathrm{ADS}^{\pi K}$ &= & $\phantom{+}0.19$ & $\pm$ & $0.19$ & $\pm$ & $0.01$, \\
$\mathcal{R}_\mathrm{ADS}^{\pi K}$ &= & $\phantom{+}0.080$ & $\pm$ & $0.015$ & $\pm$ & $0.002$, \\
$\mathcal{A}_\mathrm{ADS}^{\pi K\pi\pi}$ &= & $-0.01$ & $\pm$ & $0.24$ & $\pm$ & $0.01$, \\
$\mathcal{R}_\mathrm{ADS}^{\pi K\pi\pi}$ &= & $\phantom{+}0.073$ & $\pm$ & $0.018$ & $\pm$ & $0.002$. \\
\end{tabular}
\end{center}

All \CP\ asymmetries are compatible with zero to within two standard deviations. The values of the GLW asymmetries and ratios are found to be consistent between the two modes, within 0.8 and 1.8 standard deviations, respectively. The results for $D \to \pip\pim\pip\pim$ are in agreement with these values, after correcting for the known \CP-even content of this state.
The same observables determined for \Bs\ decays are compatible with the \CP-conserving hypothesis. Results for $\Bs$ decays can be found in Appendix~\ref{sec:appendixa}, together with the results for all observables separated between the Run 1 and Run 2 data sets, and full correlation matrices.

The statistical significances of the signal yields in  
the previously unobserved channels are calculated using Wilks' theorem~\cite{Wilks:1938dza}. The likelihood profiles are convolved with a Gaussian function with standard deviation equal to the systematic uncertainties on the yields. This procedure yields a significance of 8.4$\sigma$ for the $\Bz \to D(\pi^+\pi^-\pi^+\pi^-)\Kstarz$ decay, 5.8$\sigma$ for the $\Bz \to D(\pi^+ K^-)\Kstarz$ decay and 4.4$\sigma$ for  the $\Bz \to D(\pi^+ K^- \pi^+\pi^-)\Kstarz$ decay, constituting the first observation of the first two modes, and strong evidence for the presence of the suppressed four-body ADS channel.

\begin{figure}
    \centering
    \setlength{\tabcolsep}{0pt}
    \begin{tabular}{cc}
        \includegraphics[width=0.49\textwidth]{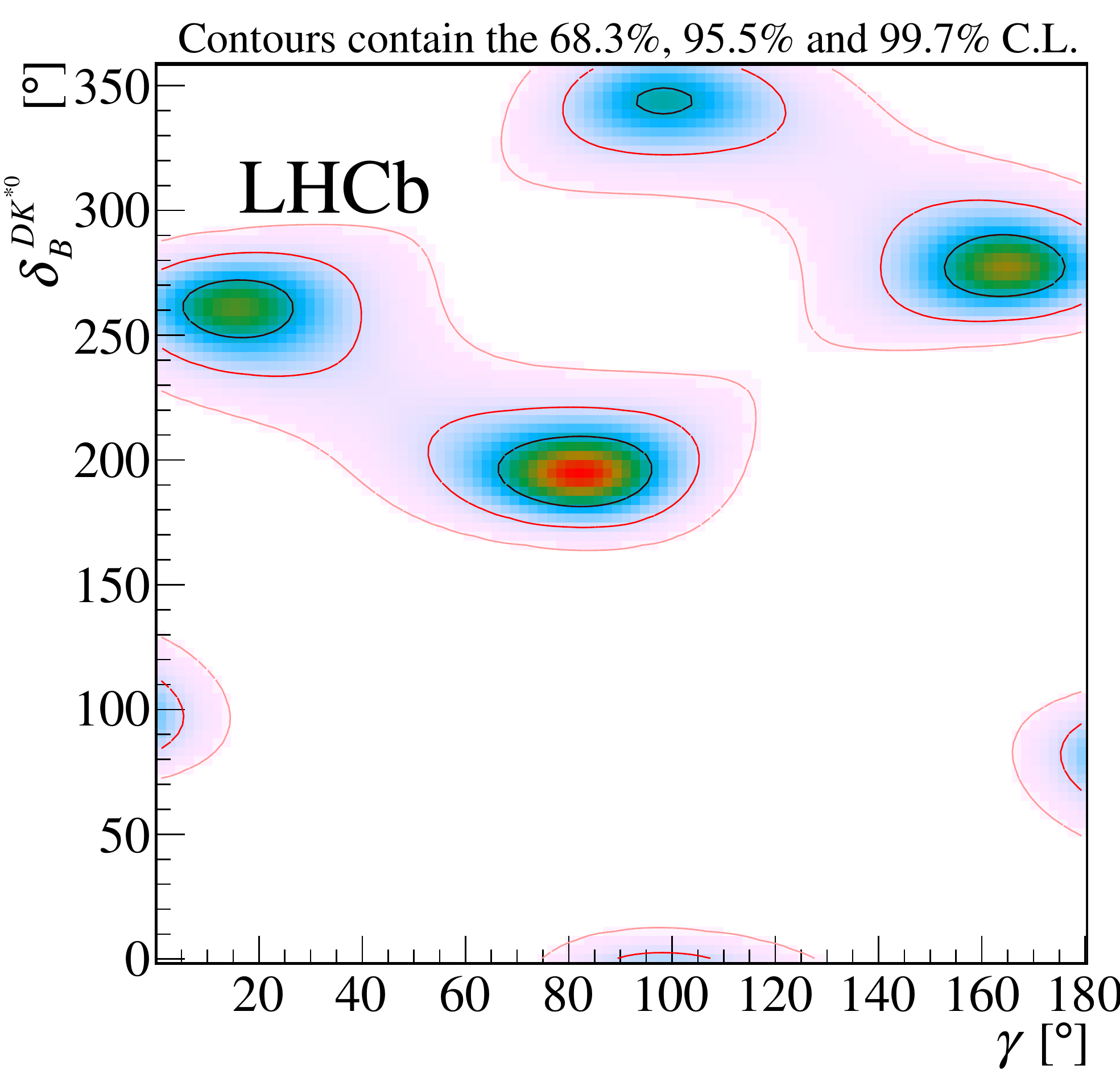} &
        \includegraphics[width=0.49\textwidth]{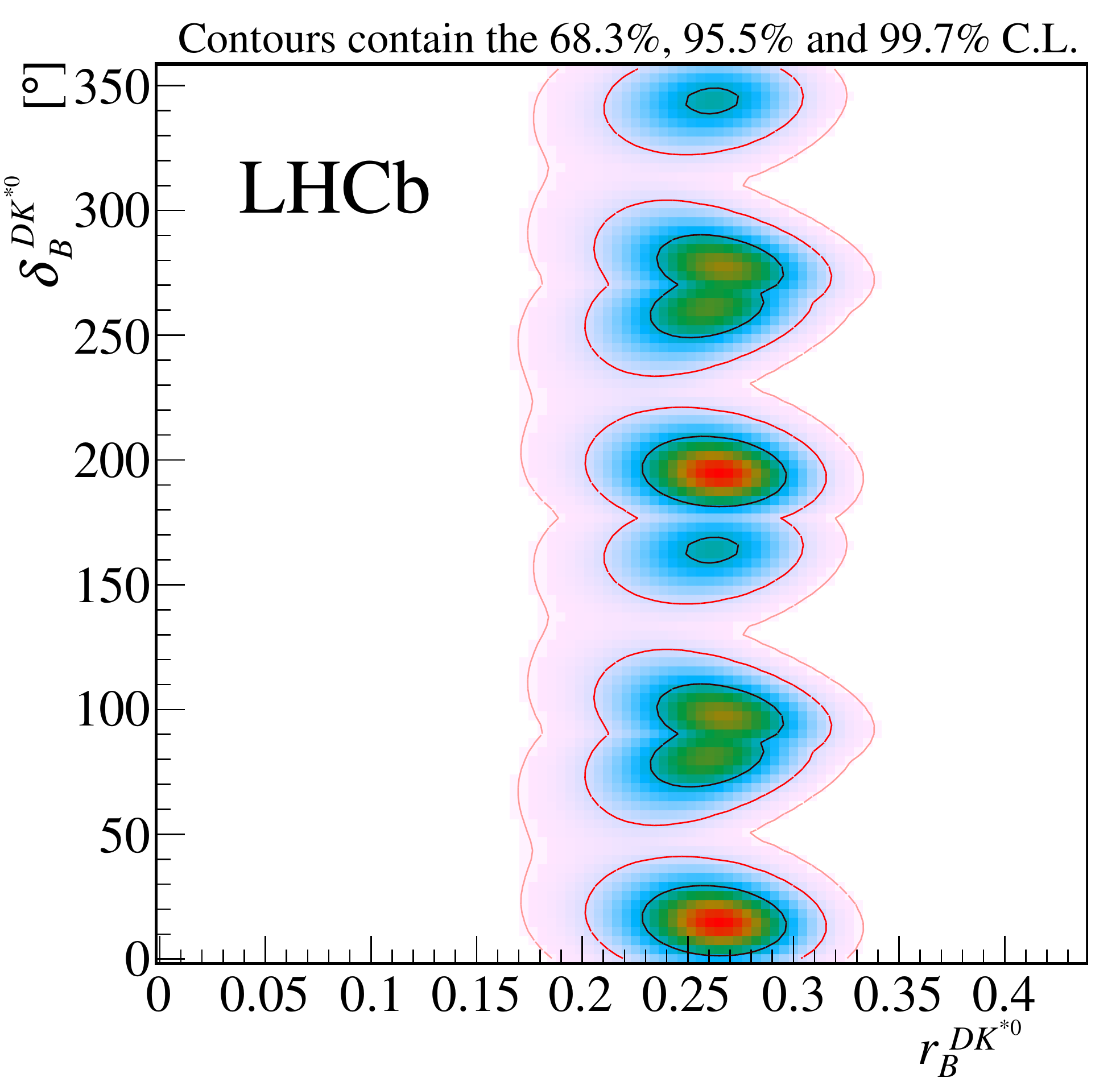} \\
    \end{tabular}
    \caption{Contour plots showing 2D scans of (left) $\delta_B^{D\!\Kstarz}$ versus $\gamma$ and (right) $\delta_B^{D\!\Kstarz}$ versus $r_B^{D\!\Kstarz}$. The lines represent the $\Delta\chisq =$\ 2.30, 6.18 and 11.8 contours, corresponding to 68.6\%, 95.5\% and 99.7\% confidence levels (C.L.), respectively.}
\label{fig:gammadini}
\end{figure}

The results are interpreted in terms of the underlying physics parameters $\gamma$, $r_B^{D\!\Kstarz}$ and $\delta_B^{D\!\Kstarz}$ by performing a global \chisq\ minimisation. The minimised \chisqndf\ is equal to 7.1/9.
A scan of physics parameters is performed for a range of values and the difference in \chisq\ between the parameter scan and the global minimum, $\Delta\chisq$, is evaluated. The confidence level for any pair of parameters is calculated assuming that these are normally distributed, which allows the $\Delta\chisq = 2.30,\ 6.18,\ 11.8$ contours to be drawn, corresponding to 68.6\%, 95.5\%, 99.7\% confidence levels, respectively. These contours are shown in Fig.~\ref{fig:gammadini}. 
As expected, there is a degeneracy in the $(\gamma, \delta_B^{D\!\Kstarz})$ plane. Four favoured solutions can be seen, two of which are compatible with the existing \lhcb\ determination of $\gamma$~\cite{LHCb-CONF-2018-002,LHCb-PAPER-2016-032}, which is dominated by results obtained from $\Bp \to D \Kp$ processes,
which have values of $r_B^{DK^-}$ and $\delta_B^{DK^-}$ different from $r_B^{D\!\Kstarz}$ and $\delta_B^{D\!\Kstarz}$. The degeneracy of the solutions can be broken by combining these results with those using other $D$ decay modes, specifically the $D \to \KS\pip\pim$ decay. The value of $r_B^{D\!\Kstarz}$ is determined to be $0.265 \pm 0.023$. In accordance with expectation, this is almost a factor of three larger than the corresponding parameter in $\Bp \to D\Kp$ decays~\cite{LHCb-CONF-2018-002,LHCb-PAPER-2016-032}. This measurement is consistent with, and more accurate than, the previous measurement by \lhcb in Ref.~\cite{LHCb-PAPER-2014-028}.

\FloatBarrier\section{Conclusion}
\label{sec:conclusion}

Measurements of \CP observables in $\Bz \to D\Kstarz$ decays with the $D$ meson decaying to $\Kp \pim$, $\pip \Km$, $\Kp \Km$ and $\pip \pim$ are performed using \lhcb data collected in 2011, 2012, 2015 and 2016.  
The results, benefitting from the increased data sample and improved analysis methods, supersede those of the  previous study~\cite{LHCb-PAPER-2014-028}.  Measurements with $D$ mesons reconstructed in the $\Kp \pim \pip \pim$, $\pip \Km \pip \pim$ and $\pip \pim \pip \pim$ final states are presented for the first time.  First observations are obtained for the suppressed ADS mode $\Bz \to D(\pip \Km) \Kstarz$ and the mode $\Bz \to D(\pip\pim\pip\pim)\Kstarz$.

The observables are interpreted in terms of the weak phase $\gamma$ and associated parameters, and are found to be compatible with the previous \lhcb\ results~\cite{LHCb-CONF-2018-002,LHCb-PAPER-2016-032}, which are dominated by measurements of $\Bp \to D \Kp$ processes. The amplitude ratio $r_B^{D\!\Kstarz}$ is determined to be equal to $0.265 \pm 0.023$ at a confidence level of 68.3\%.
These results can be combined with those from other modes in $\Bz \to D \Kstarz$ decays to provide powerful constraints on $\gamma$. This can be compared to results obtained from studies of other processes.

\section*{Acknowledgements}
%
%
\noindent We express our gratitude to our colleagues in the CERN
accelerator departments for the excellent performance of the LHC. We
thank the technical and administrative staff at the LHCb
institutes.
We acknowledge support from CERN and from the national agencies:
CAPES, CNPq, FAPERJ and FINEP (Brazil); 
MOST and NSFC (China); 
CNRS/IN2P3 (France); 
BMBF, DFG and MPG (Germany); 
INFN (Italy); 
NWO (Netherlands); 
MNiSW and NCN (Poland); 
MEN/IFA (Romania); 
MSHE (Russia); 
MinECo (Spain); 
SNSF and SER (Switzerland); 
NASU (Ukraine); 
STFC (United Kingdom); 
DOE NP and NSF (USA).
We acknowledge the computing resources that are provided by CERN, IN2P3
(France), KIT and DESY (Germany), INFN (Italy), SURF (Netherlands),
PIC (Spain), GridPP (United Kingdom), RRCKI and Yandex
LLC (Russia), CSCS (Switzerland), IFIN-HH (Romania), CBPF (Brazil),
PL-GRID (Poland) and OSC (USA).
We are indebted to the communities behind the multiple open-source
software packages on which we depend.
Individual groups or members have received support from
AvH Foundation (Germany);
EPLANET, Marie Sk\l{}odowska-Curie Actions and ERC (European Union);
ANR, Labex P2IO and OCEVU, and R\'{e}gion Auvergne-Rh\^{o}ne-Alpes (France);
Key Research Program of Frontier Sciences of CAS, CAS PIFI, and the Thousand Talents Program (China);
RFBR, RSF and Yandex LLC (Russia);
GVA, XuntaGal and GENCAT (Spain);
the Royal Society
and the Leverhulme Trust (United Kingdom).

\clearpage

{\noindent\normalfont\bfseries\Large Appendices}

\appendix
\FloatBarrier\section{Additional results}
\label{sec:appendixa}

Observables for $\Bsb \to D\Kstarz$ decays are defined analogously to those for $\Bz \to D\Kstarz$ decays in Eqs.~\ref{eq:A_hh},~\ref{eq:rhhcp} and~\ref{eq:A_ADS}.
The measured observables are 

\begin{center}
\setlength{\tabcolsep}{2pt}
\begin{tabular}{rclclcl}
$\mathcal{A}_{s,\rm ADS}^{\pi K}$ &= & $\phantom{+}0.006$ & $\pm$ & $0.017$ & $\pm$ & $0.012$, \\
$\mathcal{A}_{s,\rm ADS}^{\pi K\pi\pi}$ &= & $-0.007$ & $\pm$ & $0.021$ & $\pm$ & $0.013$, \\
$\mathcal{A}_{s,\CP}^{KK}$ &= & $\phantom{+}0.06$ & $\pm$ & $0.05$ & $\pm$ & $0.01$, \\
$\mathcal{A}_{s,\CP}^{\pi\pi}$ &= & $-0.11$ & $\pm$ & $0.09$ & $\pm$ & $0.01$, \\
$\mathcal{R}_{s,\CP}^{KK}$ &= & $\phantom{+}1.06$ & $\pm$ & $0.06$ & $\pm$ & $0.02$, \\
$\mathcal{R}_{s,\CP}^{\pi\pi}$ &= & $\phantom{+}1.05$ & $\pm$ & $0.10$ & $\pm$ & $0.02$, \\
$\mathcal{A}_{s,\CP}^{4\pi}$ &= & $\phantom{+}0.12$ & $\pm$ & $0.08$ & $\pm$ & $0.02$, \\
$\mathcal{R}_{s,\CP}^{4\pi}$ &= & $\phantom{+}0.964$ & $\pm$ & $0.086$ & $\pm$ & $0.031$. \\
\end{tabular}
\end{center}

The $\Bz$ and $\Bs$ observables are also measured separately for Run 1 and Run 2; these measurements are presented in Table~\ref{tab:results_split}. The correlation matrices for the principal observables are given in Tables~\ref{tab:correlation},~\ref{tab:correlation_1v1} and~\ref{tab:correlation_2v2} for the combined, Run 1, and Run 2 results, respectively. Table~\ref{tab:correlation_1v2} gives the correlations between the Run 1 and Run 2 results.

\begin{table}
\setlength{\tabcolsep}{2pt}
\centering
    \caption{Measured observables split by LHC running period. Observables relating to $\Bz \to D(\pip\pim\pip\pim)\Kstarz$ decays are not presented for Run 1, as this decay channel was not selected in the Run 1 data.}
\begin{tabular}{l|lclcl|lclcl}
& \multicolumn{5}{c|}{Run 1} & \multicolumn{5}{c}{Run 2}\\
\midrule
$\mathcal{A}_{\CP}^{KK}$ & $-0.19$ & $\pm$ & $0.16$ & $\pm$ & $0.01$ & $\phantom{+}0.05$ & $\pm$ & $0.13$ & $\pm$ & $0.02$ \\
$\mathcal{A}_{\CP}^{\pi\pi}$ & $-0.06$ & $\pm$ & $0.23$ & $\pm$ & $0.01$ & $-0.26$ & $\pm$ & $0.18$ & $\pm$ & $0.01$ \\
$\mathcal{R}_{\CP}^{KK}$ & $\phantom{+}0.93$ & $\pm$ & $0.15$ & $\pm$ & $0.02$ & $\phantom{+}0.91$ & $\pm$ & $0.13$ & $\pm$ & $0.02$ \\
$\mathcal{R}_{\CP}^{\pi\pi}$ & $\phantom{+}1.39$ & $\pm$ & $0.33$ & $\pm$ & $0.04$ & $\phantom{+}1.27$ & $\pm$ & $0.24$ & $\pm$ & $0.03$ \\
$\mathcal{A}_{\CP}^{4\pi}$ & \multicolumn{5}{c|}{\textemdash} & $-0.03$ & $\pm$ & $0.15$ & $\pm$ & $0.01$ \\
$\mathcal{R}_{\CP}^{4\pi}$ & \multicolumn{5}{c|}{\textemdash} & $\phantom{+}1.01$ & $\pm$ & $0.16$ & $\pm$ & $0.04$ \\
$\mathcal{R}_+^{\pi K}$ & $\phantom{+}0.045$ & $\pm$ & $0.032$ & $\pm$ & $0.003$ & $\phantom{+}0.076$ & $\pm$ & $0.027$ & $\pm$ & $0.003$ \\
$\mathcal{R}_-^{\pi K}$ & $\phantom{+}0.120$ & $\pm$ & $0.035$ & $\pm$ & $0.003$ & $\phantom{+}0.080$ & $\pm$ & $0.025$ & $\pm$ & $0.003$ \\
$\mathcal{R}_+^{\pi K\pi\pi}$ & $\phantom{+}0.12$ & $\pm$ & $0.04$ & $\pm$ & $0.00$ & $\phantom{+}0.047$ & $\pm$ & $0.031$ & $\pm$ & $0.003$ \\
    $\mathcal{R}_-^{\pi K\pi\pi}$ & $\phantom{+}0.099$ & $\pm$ & $0.043$ & $\pm$ & $0.004\phantom{+}$ & $\phantom{+}0.056$ & $\pm$ & $0.029$ & $\pm$ & $0.003$ \\
$\mathcal{A}_{\rm ADS}^{K\pi}$ & $\phantom{+}0.03$ & $\pm$ & $0.04$ & $\pm$ & $0.01$ & $\phantom{+}0.055$ & $\pm$ & $0.035$ & $\pm$ & $0.016$ \\
$\mathcal{A}_{\rm ADS}^{K\pi\pi\pi}$ & $\phantom{+}0.03$ & $\pm$ & $0.05$ & $\pm$ & $0.01$ & $\phantom{+}0.042$ & $\pm$ & $0.040$ & $\pm$ & $0.016$ \\
$\mathcal{A}_{s,\rm ADS}^{\pi K}$ & $\phantom{+}0.011$ & $\pm$ & $0.027$ & $\pm$ & $0.010$ & $\phantom{+}0.002$ & $\pm$ & $0.022$ & $\pm$ & $0.019$ \\
$\mathcal{A}_{s,\rm ADS}^{\pi K\pi\pi}$ & $-0.042$ & $\pm$ & $0.035$ & $\pm$ & $0.012$ & $\phantom{+}0.012$ & $\pm$ & $0.026$ & $\pm$ & $0.020$ \\
$\mathcal{A}_{s,\CP}^{KK}$ & $-0.03$ & $\pm$ & $0.08$ & $\pm$ & $0.01$ & $\phantom{+}0.13$ & $\pm$ & $0.07$ & $\pm$ & $0.02$ \\
$\mathcal{A}_{s,\CP}^{\pi\pi}$ & $-0.06$ & $\pm$ & $0.17$ & $\pm$ & $0.01$ & $-0.13$ & $\pm$ & $0.11$ & $\pm$ & $0.02$ \\
$\mathcal{R}_{s,\CP}^{KK}$ & $\phantom{+}1.14$ & $\pm$ & $0.09$ & $\pm$ & $0.03$ & $\phantom{+}1.01$ & $\pm$ & $0.07$ & $\pm$ & $0.02$ \\
$\mathcal{R}_{s,\CP}^{\pi\pi}$ & $\phantom{+}0.83$ & $\pm$ & $0.15$ & $\pm$ & $0.02$ & $\phantom{+}1.22$ & $\pm$ & $0.14$ & $\pm$ & $0.03$ \\
$\mathcal{A}_{s,\CP}^{4\pi}$ & \multicolumn{5}{c|}{\textemdash} & $\phantom{+}0.12$ & $\pm$ & $0.08$ & $\pm$ & $0.02$ \\
$\mathcal{R}_{s,\CP}^{4\pi}$ & \multicolumn{5}{c|}{\textemdash} & $\phantom{+}0.96$ & $\pm$ & $0.09$ & $\pm$ & $0.03$ \\
\end{tabular}
\label{tab:results_split}
\end{table}

\begin{table}
\centering
\caption{Combined statistical and systematic correlation matrix for the principal observables.}
\footnotesize
\setlength{\tabcolsep}{2pt}
\begin{tabular}{l|rrrrrrrrrrrr}
& $\mathcal{A}_{\CP}^{KK}$& $\mathcal{A}_{\CP}^{\pi\pi}$& $\mathcal{R}_{\CP}^{KK}$& $\mathcal{R}_{\CP}^{\pi\pi}$& $\mathcal{A}_{\CP}^{4\pi}$& $\mathcal{R}_{\CP}^{4\pi}$& $\mathcal{R}_+^{\pi K}$& $\mathcal{R}_-^{\pi K}$& $\mathcal{R}_+^{\pi K\pi\pi}$& $\mathcal{R}_-^{\pi K\pi\pi}$& $\mathcal{A}_{\rm ADS}^{K\pi}$& $\mathcal{A}_{\rm ADS}^{K\pi\pi\pi}$ \\
\midrule

$\mathcal{A}_{CP}^{KK}$& $1.00$& $0.00$& $0.03$& $-0.01$& $0.00$& $0.00$& $0.00$& $-0.01$& $-0.01$& $-0.01$& $-0.01$& $-0.01$ \\

$\mathcal{A}_{CP}^{\pi\pi}$& $0.00$& $1.00$& $0.01$& $0.06$& $0.00$& $0.00$& $0.00$& $0.00$& $0.00$& $0.00$& $-0.01$& $-0.01$ \\

$\mathcal{R}_{CP}^{KK}$& $0.03$& $0.01$& $1.00$& $0.04$& $0.00$& $0.03$& $0.02$& $0.02$& $0.00$& $0.00$& $-0.04$& $-0.03$ \\

$\mathcal{R}_{CP}^{\pi\pi}$& $-0.01$& $0.06$& $0.04$& $1.00$& $0.00$& $0.04$& $0.01$& $0.03$& $0.02$& $0.02$& $0.03$& $0.03$ \\

$\mathcal{A}_{CP}^{4\pi}$& $0.00$& $0.00$& $0.00$& $0.00$& $1.00$& $0.01$& $0.00$& $0.00$& $0.00$& $0.00$& $0.00$& $0.00$ \\

$\mathcal{R}_{CP}^{4\pi}$& $0.00$& $0.00$& $0.03$& $0.04$& $0.01$& $1.00$& $0.01$& $0.02$& $0.02$& $0.03$& $0.02$& $0.01$ \\

$\mathcal{R}_+^{\pi K}$& $0.00$& $0.00$& $0.02$& $0.01$& $0.00$& $0.01$& $1.00$& $0.05$& $0.01$& $0.01$& $0.08$& $0.00$ \\

$\mathcal{R}_-^{\pi K}$& $-0.01$& $0.00$& $0.02$& $0.03$& $0.00$& $0.02$& $0.05$& $1.00$& $0.02$& $0.02$& $-0.08$& $0.03$ \\

$\mathcal{R}_+^{\pi K\pi\pi}$& $-0.01$& $0.00$& $0.00$& $0.02$& $0.00$& $0.02$& $0.01$& $0.02$& $1.00$& $0.06$& $0.02$& $0.11$ \\

$\mathcal{R}_-^{\pi K\pi\pi}$& $-0.01$& $0.00$& $0.00$& $0.02$& $0.00$& $0.03$& $0.01$& $0.02$& $0.06$& $1.00$& $0.03$& $-0.06$ \\

$\mathcal{A}_{\rm ADS}^{K\pi}$& $-0.01$& $-0.01$& $-0.04$& $0.03$& $0.00$& $0.02$& $0.08$& $-0.08$& $0.02$& $0.03$& $1.00$& $0.08$ \\

$\mathcal{A}_{\rm ADS}^{K\pi\pi\pi}$& $-0.01$& $-0.01$& $-0.03$& $0.03$& $0.00$& $0.01$& $0.00$& $0.03$& $0.11$& $-0.06$& $0.08$& $1.00$ \\
\end{tabular}
\label{tab:correlation}
\end{table}

\begin{table}
\centering
\caption{Combined statistical and systematic correlation matrix for the principal observables in Run 1 data only.}
\footnotesize
\setlength{\tabcolsep}{2pt}
\begin{tabular}{l|rrrrrrrrrr}
& $\mathcal{A}_{\CP,1}^{KK}$& $\mathcal{A}_{\CP,1}^{\pi\pi}$& $\mathcal{R}_{\CP,1}^{KK}$& $\mathcal{R}_{\CP,1}^{\pi\pi}$& $\mathcal{R}_{+,1}^{\pi K}$& $\mathcal{R}_{-,1}^{\pi K}$& $\mathcal{R}_{+,1}^{\pi K\pi\pi}$& $\mathcal{R}_{-,1}^{\pi K\pi\pi}$& $\mathcal{A}_{\rm ADS,1}^{K\pi}$& $\mathcal{A}_{\rm ADS,1}^{K\pi\pi\pi}$ \\
\midrule

$\mathcal{A}_{\CP,1}^{KK}$ & $1.00$ & $0.00$ & $0.08$ & $0.00$ & $0.00$ & $0.00$ & $0.00$ & $0.00$ & $-0.01$ & $-0.01$ \\

$\mathcal{A}_{\CP,1}^{\pi\pi}$ & $0.00$ & $1.00$ & $0.00$ & $0.01$ & $0.00$ & $0.00$ & $0.00$ & $0.00$ & $0.00$ & $0.00$ \\

$\mathcal{R}_{\CP,1}^{KK}$ & $0.08$ & $0.00$ & $1.00$ & $0.03$ & $0.01$ & $0.02$ & $0.00$ & $0.00$ & $-0.02$ & $-0.01$ \\

$\mathcal{R}_{\CP,1}^{\pi\pi}$ & $0.00$ & $0.01$ & $0.03$ & $1.00$ & $0.00$ & $0.03$ & $0.01$ & $0.01$ & $0.01$ & $0.01$ \\

$\mathcal{R}_{+,1}^{\pi K}$ & $0.00$ & $0.00$ & $0.01$ & $0.00$ & $1.00$ & $0.02$ & $0.00$ & $0.00$ & $0.05$ & $-0.01$ \\

$\mathcal{R}_{-,1}^{\pi K}$ & $0.00$ & $0.00$ & $0.02$ & $0.03$ & $0.02$ & $1.00$ & $0.01$ & $0.01$ & $-0.13$ & $0.01$ \\

$\mathcal{R}_{+,1}^{\pi K\pi\pi}$ & $0.00$ & $0.00$ & $0.00$ & $0.01$ & $0.00$ & $0.01$ & $1.00$ & $0.04$ & $0.01$ & $0.15$ \\

$\mathcal{R}_{-,1}^{\pi K\pi\pi}$ & $0.00$ & $0.00$ & $0.00$ & $0.01$ & $0.00$ & $0.01$ & $0.04$ & $1.00$ & $0.01$ & $-0.11$ \\

$\mathcal{A}_{\rm ADS,1}^{K\pi}$ & $-0.01$ & $0.00$ & $-0.02$ & $0.01$ & $0.05$ & $-0.13$ & $0.01$ & $0.01$ & $1.00$ & $0.02$ \\

$\mathcal{A}_{\rm ADS,1}^{K\pi\pi\pi}$ & $-0.01$ & $0.00$ & $-0.01$ & $0.01$ & $-0.01$ & $0.01$ & $0.15$ & $-0.11$ & $0.02$ & $1.00$ \\
\end{tabular}
\label{tab:correlation_1v1}
\end{table}

\begin{table}
\centering
\caption{Combined statistical and systematic correlation matrix for the principal observables in Run 2 data only.}
\footnotesize
\setlength{\tabcolsep}{2pt}
\begin{tabular}{l|rrrrrrrrrrrr}
& $\mathcal{A}_{\CP,2}^{KK}$& $\mathcal{A}_{\CP,2}^{\pi\pi}$& $\mathcal{R}_{\CP,2}^{KK}$& $\mathcal{R}_{\CP,2}^{\pi\pi}$& $\mathcal{A}_{\CP,2}^{4\pi}$& $\mathcal{R}_{\CP,2}^{4\pi}$& $\mathcal{R}_{+,2}^{\pi K}$& $\mathcal{R}_{-,2}^{\pi K}$& $\mathcal{R}_{+,2}^{\pi K\pi\pi}$& $\mathcal{R}_{-,2}^{\pi K\pi\pi}$& $\mathcal{A}_{\rm ADS,2}^{K\pi}$& $\mathcal{A}_{\rm ADS,2}^{K\pi\pi\pi}$ \\
\midrule

$\mathcal{A}_{\CP,2}^{KK}$ & $1.00$ & $-0.01$ & $-0.01$ & $0.01$ & $0.00$ & $0.01$ & $0.01$ & $0.01$ & $-0.01$ & $-0.01$ & $0.04$ & $0.03$ \\

$\mathcal{A}_{\CP,2}^{\pi\pi}$ & $-0.01$ & $1.00$ & $0.01$ & $0.08$ & $0.00$ & $0.00$ & $-0.01$ & $-0.01$ & $0.01$ & $0.01$ & $-0.02$ & $-0.01$ \\

$\mathcal{R}_{\CP,2}^{KK}$ & $-0.01$ & $0.01$ & $1.00$ & $0.03$ & $0.00$ & $0.03$ & $0.01$ & $0.01$ & $0.01$ & $0.01$ & $-0.04$ & $-0.03$ \\

$\mathcal{R}_{\CP,2}^{\pi\pi}$ & $0.01$ & $0.08$ & $0.03$ & $1.00$ & $0.00$ & $0.03$ & $0.02$ & $0.03$ & $-0.01$ & $0.00$ & $0.03$ & $0.03$ \\

$\mathcal{A}_{\CP,2}^{4\pi}$ & $0.00$ & $0.00$ & $0.00$ & $0.00$ & $1.00$ & $0.00$ & $0.00$ & $0.00$ & $0.00$ & $0.00$ & $0.00$ & $0.01$ \\

$\mathcal{R}_{\CP,2}^{4\pi}$ & $0.01$ & $0.00$ & $0.03$ & $0.03$ & $0.00$ & $1.00$ & $0.01$ & $0.01$ & $0.00$ & $0.01$ & $0.01$ & $0.01$ \\

$\mathcal{R}_{+,2}^{\pi K}$ & $0.01$ & $-0.01$ & $0.01$ & $0.02$ & $0.00$ & $0.01$ & $1.00$ & $0.05$ & $0.00$ & $0.00$ & $0.12$ & $0.02$ \\

$\mathcal{R}_{-,2}^{\pi K}$ & $0.01$ & $-0.01$ & $0.01$ & $0.03$ & $0.00$ & $0.01$ & $0.05$ & $1.00$ & $0.00$ & $0.00$ & $-0.07$ & $0.02$ \\

$\mathcal{R}_{+,2}^{\pi K\pi\pi}$ & $-0.01$ & $0.01$ & $0.01$ & $-0.01$ & $0.00$ & $0.00$ & $0.00$ & $0.00$ & $1.00$ & $0.05$ & $-0.02$ & $0.04$ \\

$\mathcal{R}_{-,2}^{\pi K\pi\pi}$ & $-0.01$ & $0.01$ & $0.01$ & $0.00$ & $0.00$ & $0.01$ & $0.00$ & $0.00$ & $0.05$ & $1.00$ & $-0.02$ & $-0.09$ \\

$\mathcal{A}_{\rm ADS,2}^{K\pi}$ & $0.04$ & $-0.02$ & $-0.04$ & $0.03$ & $0.00$ & $0.01$ & $0.12$ & $-0.07$ & $-0.02$ & $-0.02$ & $1.00$ & $0.10$ \\

$\mathcal{A}_{\rm ADS,2}^{K\pi\pi\pi}$ & $0.03$ & $-0.01$ & $-0.03$ & $0.03$ & $0.01$ & $0.01$ & $0.02$ & $0.02$ & $0.04$ & $-0.09$ & $0.10$ & $1.00$ \\
\end{tabular}
\label{tab:correlation_2v2}
\end{table}

\begin{table}
\centering
\caption{Correlation matrix for the principal observables between Run 1 and Run 2 data.}
\footnotesize
\setlength{\tabcolsep}{2pt}
\begin{tabular}{l|rrrrrrrrrrrr}
    & $\mathcal{A}_{\CP,2}^{KK}$& $\mathcal{A}_{\CP,2}^{\pi\pi}$& $\mathcal{R}_{\CP,2}^{KK}$& $\mathcal{R}_{\CP,2}^{\pi\pi}$& $\mathcal{A}_{\CP}^{4\pi}$& $\mathcal{R}_{\CP}^{4\pi}$& $\mathcal{R}_{+,2}^{\pi K}$& $\mathcal{R}_{-,2}^{\pi K}$& $\mathcal{R}_{+,2}^{\pi K\pi\pi}$& $\mathcal{R}_{-,2}^{\pi K\pi\pi}$& $\mathcal{A}_{\rm ADS,2}^{K\pi}$& $\mathcal{A}_{\rm ADS,2}^{K\pi\pi\pi}$ \\
\midrule

$\mathcal{A}_{\CP,1}^{KK}$ & $0.00$ & $0.00$ & $0.01$ & $0.00$ & $0.00$ & $0.00$ & $0.00$ & $0.00$ & $0.00$ & $0.00$ & $-0.01$ & $-0.01$ \\

$\mathcal{A}_{\CP,1}^{\pi\pi}$ & $0.00$ & $0.00$ & $0.00$ & $0.00$ & $0.00$ & $0.00$ & $0.00$ & $0.00$ & $0.00$ & $0.00$ & $-0.01$ & $0.00$ \\

$\mathcal{R}_{\CP,1}^{KK}$ & $-0.01$ & $0.01$ & $0.03$ & $0.01$ & $0.00$ & $0.02$ & $0.00$ & $-0.01$ & $0.01$ & $0.01$ & $-0.02$ & $-0.02$ \\

$\mathcal{R}_{\CP,1}^{\pi\pi}$ & $0.01$ & $-0.01$ & $0.00$ & $0.02$ & $0.00$ & $0.02$ & $0.01$ & $0.01$ & $-0.01$ & $-0.01$ & $0.03$ & $0.02$ \\

$\mathcal{R}_{+,1}^{\pi K}$ & $-0.01$ & $0.00$ & $0.01$ & $-0.01$ & $0.00$ & $0.00$ & $0.02$ & $0.01$ & $0.01$ & $0.01$ & $-0.02$ & $-0.02$ \\

$\mathcal{R}_{-,1}^{\pi K}$ & $0.01$ & $-0.01$ & $0.00$ & $0.02$ & $0.00$ & $0.01$ & $0.02$ & $0.04$ & $0.00$ & $0.00$ & $0.03$ & $0.02$ \\

$\mathcal{R}_{+,1}^{\pi K\pi\pi}$ & $0.01$ & $-0.01$ & $0.00$ & $0.01$ & $0.00$ & $0.01$ & $0.01$ & $0.01$ & $0.02$ & $0.01$ & $0.02$ & $0.02$ \\

$\mathcal{R}_{-,1}^{\pi K\pi\pi}$ & $0.01$ & $-0.01$ & $0.00$ & $0.01$ & $0.00$ & $0.01$ & $0.01$ & $0.02$ & $0.01$ & $0.03$ & $0.02$ & $0.02$ \\

$\mathcal{A}_{\rm ADS,1}^{K\pi}$ & $0.02$ & $-0.01$ & $-0.02$ & $0.02$ & $0.00$ & $0.01$ & $0.01$ & $0.02$ & $-0.01$ & $-0.01$ & $0.05$ & $0.04$ \\

$\mathcal{A}_{\rm ADS,1}^{K\pi\pi\pi}$ & $0.02$ & $-0.01$ & $-0.02$ & $0.02$ & $0.00$ & $0.01$ & $0.01$ & $0.02$ & $-0.01$ & $-0.01$ & $0.05$ & $0.04$ \\
\end{tabular}
\label{tab:correlation_1v2}
\end{table}

\FloatBarrier

\addcontentsline{toc}{section}{References}
\bibliographystyle{LHCb}
\bibliography{main,standard,LHCb-PAPER,LHCb-CONF,LHCb-DP,LHCb-TDR}

\newpage
\centerline
{\large\bf LHCb collaboration}
\begin
{flushleft}
\small
R.~Aaij$^{29}$,
C.~Abell{\'a}n~Beteta$^{46}$,
B.~Adeva$^{43}$,
M.~Adinolfi$^{50}$,
C.A.~Aidala$^{78}$,
Z.~Ajaltouni$^{7}$,
S.~Akar$^{61}$,
P.~Albicocco$^{20}$,
J.~Albrecht$^{12}$,
F.~Alessio$^{44}$
% ...

{\footnotesize \it

% ...
$ ^{7}$Universit{\'e} Clermont Auvergne, CNRS/IN2P3, LPC, Clermont-Ferrand, France\\
% ...
$ ^{12}$Fakult{\"a}t Physik, Technische Universit{\"a}t Dortmund, Dortmund, Germany\\
% ...
$ ^{20}$INFN Laboratori Nazionali di Frascati, Frascati, Italy\\
% ...
$ ^{29}$Nikhef National Institute for Subatomic Physics, Amsterdam, Netherlands\\
% ...
$ ^{43}$Instituto Galego de F{\'\i}sica de Altas Enerx{\'\i}as (IGFAE), Universidade de Santiago de Compostela, Santiago de Compostela, Spain\\
$ ^{44}$European Organization for Nuclear Research (CERN), Geneva, Switzerland\\
% ...
$ ^{46}$Physik-Institut, Universit{\"a}t Z{\"u}rich, Z{\"u}rich, Switzerland\\
% ...
$ ^{50}$H.H. Wills Physics Laboratory, University of Bristol, Bristol, United Kingdom\\
% ...
$ ^{61}$University of Cincinnati, Cincinnati, OH, United States\\
% ...
$ ^{64}$Syracuse University, Syracuse, NY, United States\\
% ...
$ ^{78}$University of Michigan, Ann Arbor, United States, associated to $^{64}$\\
\bigskip
% ...
}
\end{flushleft}

\end{document}